\documentclass[pre,twocolumn,showpacs,superscriptaddress]{revtex4-1}

\usepackage{amsmath,graphicx}

\def \dif {{d}}

\date{\today}

\begin{document}

\title{Decay of Bogoliubov excitations in one-dimensional Bose gases}

\author{Zoran Ristivojevic}
\affiliation{Laboratoire de Physique Th\'{e}orique, Universit\'{e} de Toulouse, CNRS, UPS, France}

\author{K. A. Matveev}
\affiliation{Materials Science Division, Argonne National Laboratory, Argonne, Illinois 60439, USA}

\begin{abstract} 
  We study the decay of Bogoliubov quasiparticles in one-dimensional
  Bose gases. Starting from the hydrodynamic Hamiltonian, we develop a
  microscopic theory that enables one to systematically study both the
  excitations and their decay. At zero temperature, the leading
  mechanism of decay of a quasiparticle is disintegration into three
  others.  We find that low-energy quasiparticles (phonons) decay with
  the rate that scales with the seventh power of momentum, whereas the
  rate of decay of the high-energy quasiparticles does not depend on
  momentum.  In addition, our approach allows us to study analytically
  the quasiparticle decay in the whole crossover region between the
  two limiting cases.  When applied to integrable models, including
  the Lieb-Liniger model of bosons with contact repulsion, our theory
  confirms the absence of the decay of quasiparticle excitations.  We
  account for two types of integrability-breaking perturbations that
  enable finite decay: three-body interaction between the bosons and
  two-body interaction of finite range.
\end{abstract}

\maketitle

\section{Introduction}

At low temperatures three-dimensional Bose gas undergoes Bose-Einstein
condensation, characterized by macroscopic occupation of the
zero-momentum state.  This feature enabled Bogoliubov in 1947 to
develop a mean field theory of weakly-interacting Bose gas
\cite{bogoliubov47,stringaribook}.  In this theory, the excitation
spectrum acquires the so-called Bogoliubov form:
\begin{align}\label{bog disp}
\varepsilon_q=\sqrt{v^2 q^2+\left(\frac{q^2}{2m}\right)^2}.
\end{align}
Here $v$ is the sound velocity, $m$ denotes the mass of bosonic
particles, while $q$ is the momentum. At low momenta, $q\ll mv$,
Bogoliubov quasiparticles are phonons with linear spectrum.  At high
momenta, $q\gg mv$, the quasiparticle energy (\ref{bog disp}) reproduces
the quadratic spectrum of the physical particles forming the Bose gas.

Bogoliubov's mean field approach neglects the residual interaction
between the quasiparticles.  As a result of these interactions,
quasiparticles are not entirely free and eventually decay.  In three
dimensions, the leading mechanism is the decay of a quasiparticle into
two others.  For quasiparticle excitations of low momenta, $q\ll mv$,
the decay rate at zero temperature was found in 1958 by Beliaev
\cite{beliaev58,stringaribook}.  It scales with the fifth power of the
quasiparticle momentum.

The decay of quasiparticles is reflected in the dynamic structure
factor of interacting bosons. It does not have the form of an
infinitely sharp delta function, but rather that of a peak with the
width determined by the decay rate.  Alternatively, the decay rate can
be probed by measuring the cross section for collisions of a
quasiparticle with the particles of the condensate.  The latter
technique was used recently
\cite{katz+02-beliaevdamping3d-PhysRevLett.89.220401} (see also
Ref.~\cite{hodby+01PhysRevLett.86.2196}) to confirm the predictions of
the Beliaev theory in three-dimensional Bose-Einstein condensates.

\begin{figure}
\includegraphics[width=0.7\columnwidth]{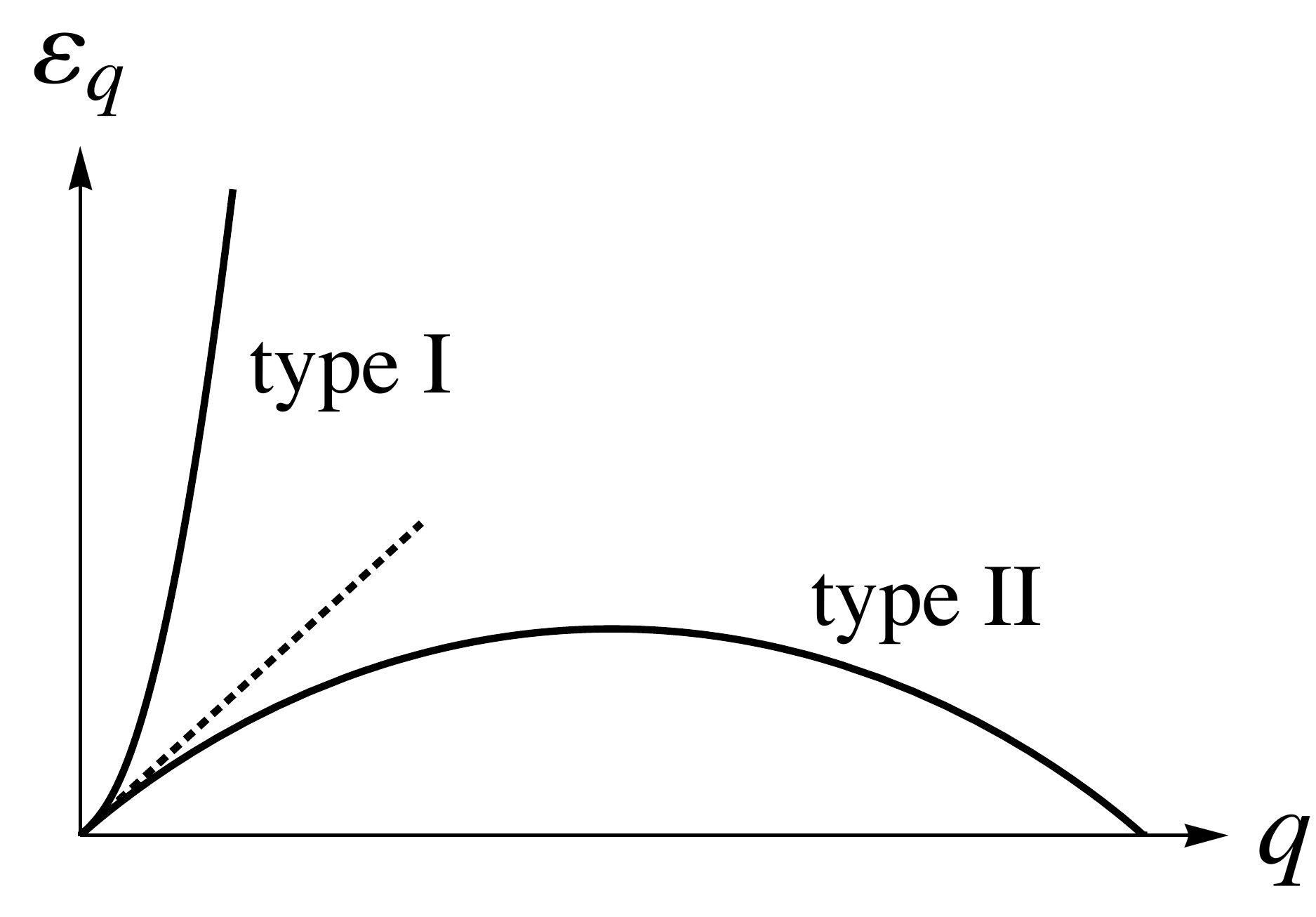}
\caption{Two branches of excitations in a one-dimensional system of bosons with contact repulsion. At small momenta the excitations on both branches are characterized by the linear spectrum, $\varepsilon_q=v |q|$, represented by the dotted line. At weak interaction, the dispersion of type I excitations deviates from linearity as $|q|^3$, while for type II as $|q|^{5/3}$. Such form of the deviation is actually true above the very small quantum crossover momentum, as we discuss further below.}\label{figLL}
\end{figure}

In contrast to the three-dimensional case, bosons in one dimension do
not condense due to the enhanced role of quantum fluctuations.
Therefore, the Bogoliubov mean-field approach cannot be applied.
Instead, Lieb and Liniger \cite{lieb1963exact} studied the model of
one-dimensional bosons with contact repulsion, which allows an exact
solution.  This enabled them to study both the ground state properties
of the system \cite{lieb1963exact} and its elementary excitations
\cite{lieb1963excitations}.  Importantly, unlike the three-dimensional
case, there are two branches of elementary excitations, see
Fig.~\ref{figLL}.  Excitation of type I behaves qualitatively similar
to the Bogoliubov mode in three dimensions, and in the limit of weak
interaction has been shown \cite{kulish+1976comparison} to have the
dispersion (\ref{bog disp}).  The second, type II excitation exists in
the limited range of momenta determined by the density and describes
the so-called dark soliton \cite{kulish+1976comparison,stringaribook}.
At the lowest momenta the two branches approach each other, having the
common linear part of the spectrum, see Fig.~\ref{figLL}.

The type II branch bends down and thus represents the lowest energy
state of the system for a given momentum.  Therefore, at zero
temperature these excitations cannot decay.  On the other hand,
momentum and energy conservation laws do not forbid the decay of the
excitation of type I.  A simple analysis shows that these excitations
still cannot decay into two others, but decay into three
quasiparticles is allowed.  In addition to momentum and energy,
integrable models possess a macroscopic number of additional conserved
quantities.  This prevents any quasiparticle decay.  On the other
hand, in practice no system is exactly integrable, and even the
smallest deviation from integrability leads to a finite decay of
quasiparticles. 

Decay of quasiparticle excitations in one-dimensional quantum liquids
is a subject of great current interest
\cite{khodas+07PhysRevB.76.155402, gangardt+10PhysRevLett.104.190402,
  tan+10relaxation, karzig+10PhysRevLett.105.226407,
  micklitz2011thermalization, ristivojevic_relaxation_2013,
  Lin+13PhysRevLett.110.016401, matveevfurusaki13,
  ristivojevic_decay_2014,protopopov2014relaxation,
  PhysRevB.91.195110}.  In this paper we study the decay of Bogoliubov
quasiparticles in a system of weakly-interacting bosons.  In the limit
of high energy of the initial quasiparticle, $q\gg mv$, this problem
was addressed in Ref.~\cite{tan+10relaxation}.  The integrability of
the Lieb-Liniger model was broken by the addition of weak three-body
interaction \cite{muryshev_dynamics_2002,mazets_breakdown_2008}.  It
was shown that this perturbation leads to a finite decay rate that
does not depend on the quasiparticle momentum.  Unlike
Ref.~\cite{tan+10relaxation}, our theory enables one to study
analytically the decay of quasiparticles of arbitrary momenta.
Furthermore, in addition to the effects of the three-body interaction,
we study another integrability-breaking perturbation, which accounts
for a finite range of two-body interaction.  This complementary term
turns out to be an important factor that also affects the decay rate.
A summary of our results for the decay of quasiparticles of small
momenta, $q\ll mv$, has been reported in
Ref.~\cite{ristivojevic_decay_2014}, where we relied on certain
phenomenological properties of one-dimensional quantum liquids.  The
approach of the present paper is fully microscopic and enables us to
find the decay rate of Bogoliubov quasiparticles in the whole range of
momenta.  In the cases $q\gg mv$ and $q\ll mv$, we recover the results
of Refs.~\cite{tan+10relaxation} and \cite{ristivojevic_decay_2014}.

The description of the excitation spectrum of weakly interacting Bose
gas in terms of Bogoliubov quasiparticles and dark solitons is
applicable only at sufficiently high momenta, $q\gg q^*$, where
$q^*\sim (mv)^{3/2}(\hbar n_0)^{-1/2}\ll mv$ and $n_0$ is the mean
particle density \cite{khodas_photosolitonic_2008,
  imambekov+12RevModPhys.84.1253, pustilnik+14PhysRevB.89.100504}.
Below the momentum scale $q^*$ the excitations are effective fermions
\cite{rozhkov2005fermionic, imambekov+12RevModPhys.84.1253}, with type
I and type II branches corresponding to quasiparticles and quasiholes,
respectively.  At zero temperature, fermionic quasiparticles decay
with the rate that scales as the eighth power of momentum
\cite{khodas+07PhysRevB.76.155402, matveevfurusaki13}.  We apply the
results of Ref.~\cite{matveevfurusaki13} to evaluate this rate in our
system, thereby presenting a complete theory of the decay of type I
excitations at zero temperature.

The paper is organized as follows.  In Sec.~\ref{sec:model} we present
the hydrodynamic description of the system of weakly interacting
bosons.  We discuss various terms in the gradient expansion and split
the Hamiltonian into a harmonic part describing the Bogoliubov
quasiparticles and the anharmonic part that accounts for their
interactions.  In Sec.~\ref{sec:amplitude} we calculate and analyze
the scattering matrix describing the decay of Bogoliubov
quasiparticles with momenta $q\gg q^*$.  The rate of decay is
evaluated in Sec.~\ref{sec:decayrate}.  In Sec.~\ref{sec:fermions} we
obtain the rate of decay of fermionic quasiparticles at momenta $q\ll
q^*$.  We discuss our results in Sec.~\ref{sec:discussions}.  Some
technical details of our work are presented in the appendices.

\section{Hamiltonian of weakly interacting bosons}\label{sec:model}

\subsection{Microscopic model}

In the representation of second quantization,
the system of interacting bosons in one dimension is described by the Hamiltonian 
\begin{equation}
\label{H}
H=H_{\text{kin}}+H_{\text{int}},
\end{equation}
where
\begin{gather}\label{Hkin1}
H_{\text{kin}}=\frac{\hbar^2}{2m} \int\dif x (\nabla\Psi^\dagger)(\nabla\Psi),\\
\label{Hint1}
H_{\text{int}}=\frac{1}{2}\int\dif x\dif x'\,g(x-x')n(x)n(x').
\end{gather}
Here Eq.~(\ref{Hkin1}) is the kinetic energy, while Eq.~(\ref{Hint1})
describes the interaction between the bosons. By $\Psi(x)$ and
$\Psi^\dagger(x)$ we denote the bosonic single particle field
operators that satisfy the standard commutation relation
$[\Psi(x),\Psi^\dagger(x')]=\delta(x-x')$. The mass of bosonic particles
is $m$. The repulsive two-particle interaction in Eq.~(\ref{Hint1}) is
described by the short-ranged function $g(x)$, while
$n=\Psi^\dagger\Psi$ denotes the density of particles. In the following we consider the case of weak interaction. This regime is defined by the condition
\begin{align}\label{cweak}
\int\dif x g(x)\ll \frac{\hbar^2 n_0}{m},
\end{align}
where $n_0$ denotes the mean density.

The Hamiltonian $H$ provides a microscopic description for an
arbitrary system of bosons in one dimension interacting via a pairwise
interaction.  In some special cases Eqs.~(\ref{H})--(\ref{Hint1})
describe the so-called integrable models. Throughout this paper, we
will be particularly interested in the Lieb-Liniger model, which is
defined by the contact interaction $g(x)=g \delta(x)$. The
integrability of this model allows an exact solution by means of the
Bethe ansatz technique \cite{lieb1963exact,lieb1963excitations}.  On
the other hand, because of integrability there is no decay of
quasiparticle excitations in the this model.  In this paper we
consider leading corrections to the Lieb-Liniger model that break the
integrability and thus ensure the decay of quasiparticles.  Since
there is no well established way to develop perturbation theory
starting with Bethe ansatz, here we develop an alternative theoretical
description.  It is based on the microscopic hydrodynamic approach
that enables us to study both the excitations and their decay.  Unlike
Bethe ansatz, this approach is limited to weak interactions, but it
has the advantage that its applicability is not limited to integrable
models.

Experimentally, the system of one-dimensional bosons can nowadays be
routinely realized with cold atomic gases
\cite{blochRevModPhys.80.885}. Starting from the three-dimensional
system of bosons, one applies an external potential to confine the
particle motion to one dimension.  At energies smaller than the
inter-subband spacing of the confining potential, one effectively
obtains a one-dimensional system of interacting bosons.  In such
situations, making use of the Hamiltonian in the form
(\ref{H})--(\ref{Hint1}) to describe the system is a priori not
justified.  Instead, one must carefully derive the corresponding
one-dimensional model.  For a typical experimental situation of
bosonic atoms in a harmonic confining potential interacting via a
short-range potential \cite{olshanii1998}, the effective
one-dimensional model is derived in several papers
\cite{muryshev_dynamics_2002,mazets_breakdown_2008,tan+10relaxation}. The
kinetic energy in the effective model of bosons is still described by
Eq.~(\ref{Hkin1}).  However, the interaction term takes a more
complicated form
\begin{align}\label{Hint}
H_\text{int}'=\frac{1}{2}\int\dif x\dif x'\,g(x-x')n(x)n(x') -\frac{\hbar^2}{m}\alpha\int\dif x\,n^3.
\end{align}
In Refs.~\cite{muryshev_dynamics_2002, mazets_breakdown_2008,
  tan+10relaxation}, the two-body interaction in Eq.~(\ref{Hint}) was
found to be of the contact type, $g(x)=g\delta(x)$.  In comparison to
Eq.~(\ref{Hint1}), the last term in Eq.~(\ref{Hint}) is new and has
the meaning of effective three-body interaction.  It was obtained
\cite{muryshev_dynamics_2002, mazets_breakdown_2008, tan+10relaxation}
by accounting for the effect of virtual transitions of bosons into
higher radial modes.

An important property of the last term in the interaction Hamiltonian
(\ref{Hint}) is that it breaks the integrability of the Lieb-Liniger
model, and thus enables the decay of quasiparticles.  In addition, we
modify the interaction Hamiltonian by assuming that the two-body
interaction potential $g(x)$ has finite width, which amounts to adding
another integrability-breaking perturbation.  In the following we refer to
$H$ as defined by $H=H_{\text{kin}}+H_{\text{int}}'$ and treat both
perturbations on equal footing.

\subsection{The density-phase representation}

\label{density-phase}

The Hamiltonian of the system of weakly interacting bosons, given by
Eqs.~(\ref{H}), (\ref{Hkin1}), and (\ref{Hint}), is expressed in terms
of the bosonic field operators $\Psi(x)$ and $\Psi^\dagger(x)$.  For our purposes it is
convenient to apply the hydrodynamic approach
\cite{popov72,haldane81prl,cazalilla+2011RMP}, in which the field
operators are expressed in terms of the particle density $n(x)$ and
its conjugate field $\theta(x)$ that can be thought of as the
superfluid phase.  In the regime of weak interaction the resulting
Hamiltonian is then naturally expressed as a sum of the contribution
$H_0$ that is quadratic in the new fields and the higher-order
perturbations $V_3$, $V_4$, etc.  In this representation $H_0$
naturally accounts for the Bogoliubov quasiparticles, while the
perturbations describe the interactions between quasiparticles that
enable their decay.

We start by expressing the bosonic field operators in terms of the
density and phase fields using the the so-called Madelung
representation \cite{popov72,haldane81prl}
\begin{align}\label{madelung}
\Psi=e^{-i\theta}\sqrt n,\quad \Psi^\dagger=\sqrt n\,e^{i\theta}.
\end{align}
The operators $\Psi(x)$ and $\Psi^\dagger(x)$ expressed in this fashion have the usual
bosonic commutation relations provided
$[n(x),\theta(x')]=-i\delta(x-x')$.  Substituting Eq.~(\ref{madelung})
into the kinetic energy (\ref{Hkin1}) of the Hamiltonian, one obtains
\cite{popov72}
\begin{align}\label{Hkin}
H_{\mathrm{kin}}=\frac{\hbar^2}{2m}\int\dif x\left[n(\nabla\theta)^2+\frac{(\nabla n)^2}{4n}\right].
\end{align}

The next step is to express the density as
\begin{align}\label{density}
n=n_0+\frac{1}{\pi}\nabla\varphi,
\end{align}
where $n_0$ is the mean particle density and the new bosonic field
$\varphi$ satisfies the commutation relation
\begin{equation}
  \label{eq:commutator_phi_theta}
  [\nabla\varphi(x),\theta(x')]=-i\pi\delta(x-x').  
\end{equation}
The hydrodynamic approach is applicable as long as the length scale
associated with the density fluctuations is large compared with the
distance between particles $n_0^{-1}$.  In this regime the density
fluctuations are small, $|\nabla\varphi|\ll n_0$, and the square root
in Eq.~(\ref{madelung}) is real.

We now take advantage of the smallness of $|\nabla\varphi|/n_0$ and
expand the Hamiltonian in powers of bosonic fields $\varphi$ and
$\theta$.  The expansion starts with quadratic contributions.  The
standard Luttinger liquid form
\begin{align}\label{HLL}
H_{LL}=\int\dif x \left[\frac{\hbar^2n_0}{2m}(\nabla\theta)^2 +\frac{g}{2\pi^2}(\nabla\varphi)^2\right]
\end{align}
is obtained from the first term in the kinetic energy (\ref{Hkin}) and
the first term in Eq.~(\ref{Hint}).  Here $g=g_0$, where
$g_q=\int\dif x e^{-iq x/\hbar}g(x)$ denotes the Fourier
transform of the interaction potential.

Apart from Eq.~(\ref{HLL}), there are a number of additional
quadratic terms in the Hamiltonian.  First, the three-body interaction
in Eq.~(\ref{Hint}) upon substitution of Eq.~(\ref{density}) generates
the contribution
\begin{align}\label{H03}
-\frac{3\alpha\hbar^2n_0}{\pi^2 m}\int\dif x (\nabla\varphi)^2.
\end{align}
Second, the so-called quantum pressure, given by the second term in
Eq.~(\ref{Hkin}), and the two-particle interaction term in
Eq.~(\ref{Hint}) give rise to
\begin{align}\label{H2kinint}
\frac{\chi^2\hbar^2}{8\pi^2 m n_0}\int\dif x(\nabla^2\varphi)^2,
\end{align} 
where
\begin{align}\label{chi2}
\chi^2=1+2mn_0 \frac{\dif^2 g_q}{\dif q^2}\bigg{|}_{q=0}.
\end{align}
For contact interaction, $g_q=\mathrm{const}$, i.e.,
$\chi^2=1$. In this special case the two-particle interaction does not contribute to Eq.~(\ref{H2kinint}). Finally, for noncontact interactions the first term in
Eq.~(\ref{Hint}) generates contributions proportional to
$(\nabla^3\varphi)^2$, $(\nabla^4\varphi)^2$, etc.  Such contributions become important only at very high momenta and therefore will be neglected.

Collecting the terms of Eqs.~(\ref{HLL}), (\ref{H03}), and (\ref{H2kinint}), we obtain the quadratic Hamiltonian
\begin{align}\label{H0}
H_0=\frac{\hbar v}{2\pi}\int\dif x \left\{K(\nabla\theta)^2  +\frac{1}{K}\left[(\nabla\varphi)^2 +\frac{2\chi^2\hbar^2}{{q}_0^2}(\nabla^2\varphi)^2\right] \right\}.
\end{align}
In Eq.~(\ref{H0}), the sound velocity $v$ satisfies
\begin{align}\label{v}
v^2=\frac{gn_0}{m}-\frac{6\alpha\hbar^2 n_0^2}{m^2},
\end{align}
the crossover momentum ${q}_0$ is introduced as
\begin{gather}
\label{q0}
{q}_0=\sqrt{8}mv,
\end{gather}
while the Luttinger liquid parameter is defined as
\begin{align}\label{K}
K=\frac{\pi\hbar n_0}{mv}.
\end{align}
The regime of weak interactions considered in this paper corresponds
to $K\gg1$, cf. Eq.~(\ref{cweak}).

The strength of the three-particle interaction is quantified by the
dimensionless coupling constant $\alpha$ [see Eq.~(\ref{Hint})].
In this paper we will require this perturbation to have only a weak
effect on the physical properties of the Bose gas.  It is instructive
to consider the effect of the three-particle interaction on the sound
velocity.  From Eq.~(\ref{v}) we conclude that the correction to $v$
is small provided
\begin{align}\label{A}
A=K^2\alpha \ll 1.
\end{align}
Since $K\gg1$, this condition is more restrictive than the naive
expectation $\alpha\ll1$.  We will see below that other physical
quantities of interest are also controlled by the parameter $A$ rather
than $\alpha$.

In addition to $H_0$, the original hydrodynamic Hamiltonian contains a
number of higher order in $\varphi$ and $\theta$ contributions that
describe the interactions between quasiparticles.  The cubic
correction to $H_0$ is
\begin{align}\label{V3general}
V_3={}&\frac{\hbar^2}{m}\int\dif x \biggl[a_1 (\nabla\varphi)(\nabla\theta)^2- \frac{a_2}{n_0^2}(\nabla^2\varphi)^2(\nabla\varphi)\notag\\
  &-\frac{\alpha}{\pi^3}(\nabla\varphi)^3 \biggr],
\end{align}
where for convenience we introduced
\begin{align}\label{a1a2}
a_1=\frac{1}{2\pi},\quad a_2=\frac{1}{8\pi^3}.
\end{align}
The first term in Eq.~(\ref{V3general}) arises from the first term in
the kinetic energy (\ref{Hkin}). The second term in
Eq.~(\ref{V3general}) emerges from the expansion of the second term in
Eq.~(\ref{Hkin}).  The last term in Eq.~(\ref{V3general}) originates
from the second term in Eq.~(\ref{Hint}).

In order to evaluate the decay rate of excitations with momenta $q\sim
q_0$ one has to account for the quartic in $\varphi$ and $\theta$
contributions to the Hamiltonian.  We write the corresponding term as
\begin{align}\label{V4general}
V_4=\frac{\hbar^2}{mn_0}\int\dif x \left[\frac{{a}_3}{n_0^2} (\nabla^2\varphi)^2(\nabla\varphi)^2+\beta (\nabla\varphi)^4\right],
\end{align}
where
\begin{align}\label{a3}
{a}_3=\frac{1}{8\pi^4}, \quad \beta=0.
\end{align}
The first term in Eq.~(\ref{V4general}) appears from the expansion of
the quantum pressure term in Eq.~(\ref{Hkin}).  The second term in
$V_4$ is not generated in the formal expansion of the Hamiltonian
given by Eqs.~(\ref{Hint}) and (\ref{Hkin}).  We added it to
Eq.~(\ref{V4general}) with a formally vanishing coefficient for
completeness and future convenience (see Appendix \ref{section:lagrange}).

So far we have expanded our Hamiltonian to the fourth order in the
bosonic fields.  The terms $V_3$ and $V_4$ will be used to evaluate
the decay rate of Bogoliubov quasiparticles with momenta of order
$q_0$, where the crossover from linear to quadratic behavior of the
quasiparticle dispersion (\ref{bog disp}) occurs.  To understand why
the subsequent higher-order terms can be neglected, one can analyze
the low-energy scaling of the Hamiltonian.  Such an analysis is
performed in Appendix \ref{scaling}, where we show that our expansion
of the Hamiltonian in powers of the bosonic fields $\varphi$ and
$\theta$ is in fact expansion in small parameter $1/\sqrt K$.  In
particular, we find $V_3\propto 1/\sqrt{K}$ and $V_4\propto 1/K$.

\subsection{Normal mode expansion}
\label{normal}

Our next goal is to obtain Bogoliubov quasiparticles as normal modes
of the quadratic Hamiltonian (\ref{H0}).  To this end we express the
fields $\varphi$ and $\theta$ in terms of bosonic quasiparticle
operators $b_q$ and $b_q^\dagger$ via the relations
\begin{gather}\label{nablaphi}
\nabla\varphi(x)=\sum_q \sqrt{\frac{\pi^2 n_0}{2Lm\varepsilon_q}}|q|e^{i q x/\hbar} (b_{-q}^\dagger+b_q),\\
\label{nablatheta}
\nabla\theta(x)=\sum_q \sqrt{\frac{m \varepsilon_q}{2 L\hbar^2 n_0}}\,\text{sgn}(q)e^{i q x/\hbar} (b^\dagger_{-q}-b_{q}).
\end{gather}
Here $L$ denotes the system size.  As a result, the Hamiltonian
(\ref{H0}) takes the diagonal form
\begin{align}\label{H0diag}
H_{0}=\sum_q\varepsilon_q b_q^\dagger b_q,
\end{align}
with the excitation spectrum given by
\begin{align}\label{Eq}
\varepsilon_q=\sqrt{v^2q^2+\chi^2\left(\frac{q^2}{2m}\right)^2}.
\end{align}
For the Lieb-Liniger model, we have $\chi=1$, and the spectrum
coincides with the well known expression (\ref{bog disp}),
Ref.~\cite{kulish+1976comparison}.

Deviation of the spectrum (\ref{Eq}) from the form (\ref{bog disp})
appears in the case of nonvanishing range of interactions between the
bosons.  This deviation is most important at high momenta $q\gg q_0$,
where $\varepsilon_q \simeq \chi q^2/2m$ rather than $q^2/2m$.  The
latter expression represents the energy of a highly excited boson,
which essentially does not interact with other bosons because of its
high momentum $q$.  This physics is not captured by the hydrodynamic
theory, which is applicable only at $q\ll \hbar n_0$.

As we show in Appendix~\ref{scaling}, the anharmonic terms
(\ref{V3general}) and (\ref{V4general}) represent corrections to
the quadratic Hamiltonian $H_0$ that are small as $1/\sqrt K$ and
$1/K$, respectively.  As a result, they do not affect the excitation spectrum
significantly.  On the other hand, they represent the residual
interactions between the quasiparticles that enable finite decay rate.
Using the normal mode representation (\ref{nablaphi}) and
(\ref{nablatheta}), the cubic anharmonic term (\ref{V3general})
becomes
\begin{align}\label{V3}
V_3={}&\frac{\pi v^2}{\sqrt{8Lmn_0}} \sum_{q_1,q_2,q_3}\frac{|q_1q_2q_3|} {\sqrt{\varepsilon_{q_1}\varepsilon_{q_2}\varepsilon_{q_3}}} \delta_{q_1+q_2+q_3,0}\notag\\
&\times\biggl[\frac{1}{3}f_+\left(q_1,q_2,q_3\right) (b^\dagger_{q_1}b^\dagger_{q_2}b^\dagger_{q_3} + \text{h.c.})\notag\\ &+f_-\left(q_1,q_2,q_3\right) (b^\dagger_{q_1} b^\dagger_{q_2}b_{-q_3} + \text{h.c.}) \biggr],
\end{align}
where the dimensionless functions are
\begin{align}\label{fpm}
f_{\pm}(q_1,q_2,q_3)={}&\frac{{a}_1}{v^2}\left( \frac{\varepsilon_{q_1}\varepsilon_{q_2}} {q_1q_2} \pm \frac{\varepsilon_{q_1}\varepsilon_{q_3}} {q_1q_3} \pm\frac{\varepsilon_{q_2}\varepsilon_{q_3}} {q_2q_3}\right)\notag\\
&+\frac{8\pi^2{a}_2}{q_0^2}(q_1q_2+q_1q_3+q_2q_3) -\frac{3A}{\pi^3}.
\end{align}
Similarly, the quartic anharmonic term (\ref{V4general}) transforms to
\begin{align}\label{V4}
V_4={}&\frac{\pi^2  v^2}{4Lmn_0} \sum_{q_1,q_2,q_3,q_4}\biggl[f(q_1,q_2,q_3,q_4) \delta_{q_1+q_2+q_3+q_4,0}\notag\\ &\times\prod_{i=1}^{4}\frac{|q_i|}{\sqrt{\varepsilon_{q_i}}} (b^\dagger_{q_i}+b_{-q_i})\biggr],
\end{align}
where
\begin{align}\label{f3}
f(q_1,q_2,q_3,q_4) ={}&-\frac{4\pi^2{a}_3}{3q_0^2} (q_1q_2+ q_1q_3+q_1q_4\notag\\
&+q_2q_3+q_2q_4+q_3q_4)+B,
\end{align}
where $B=K^2\beta$.  We will now apply the results
(\ref{V3})--(\ref{f3}) to the evaluation of the decay rate of
Bogoliubov quasiparticles.

\section{Scattering matrix element}\label{sec:amplitude}

The spectrum of a Bogoliubov quasiparticle in a weakly interacting
Bose gas is given by Eq.~(\ref{Eq}).  The presence in the Hamiltonian
of weak anharmonic perturbations, such as $V_3$ and $V_4$, means that
the quasiparticles are weakly interacting.  This generally leads to
their decay.  Our goal is to study the decay of a state with a single
quasiparticle as a function of its momentum $Q$.

\begin{figure}
\includegraphics[width=0.6\columnwidth]{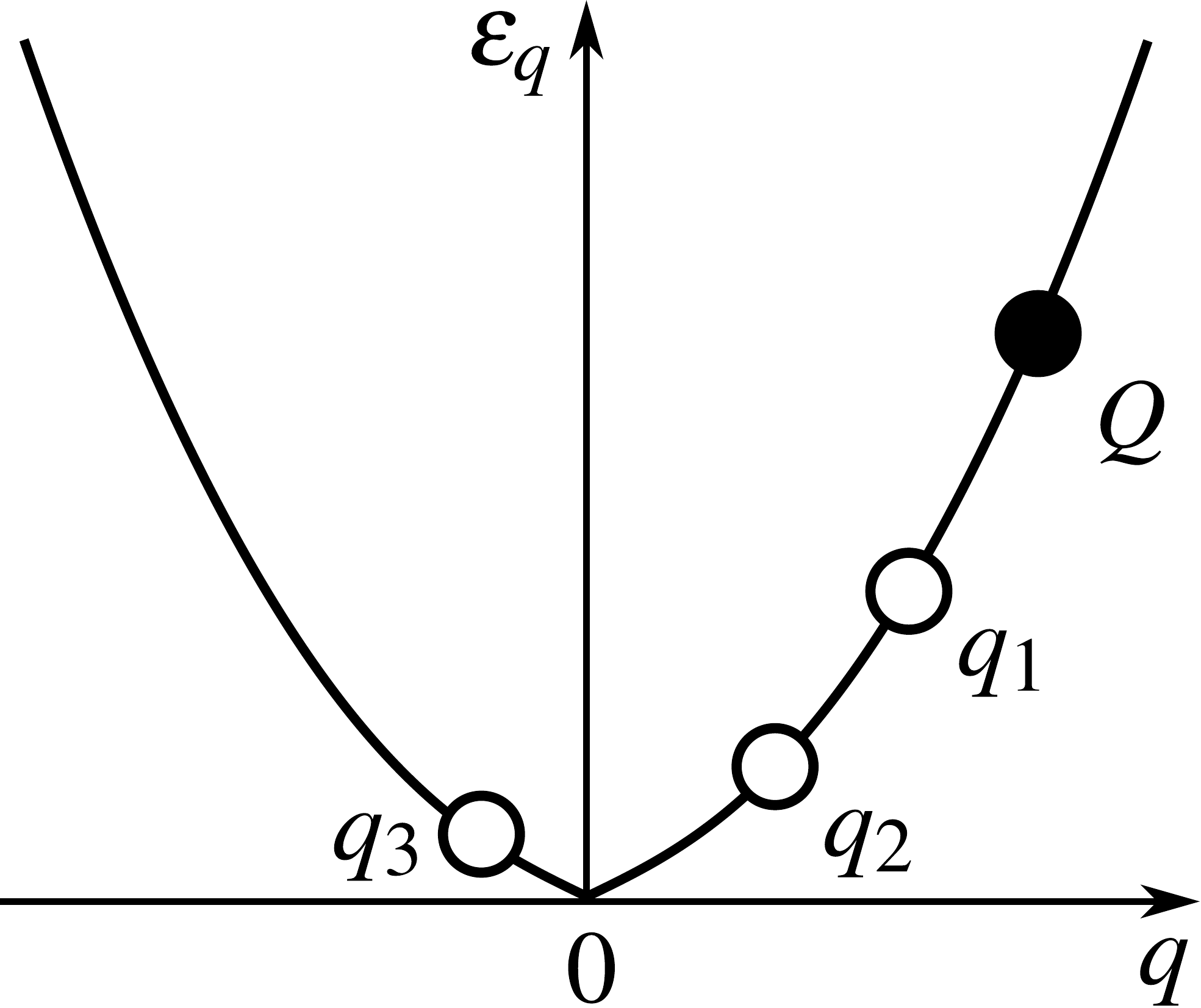}
\caption{In a one dimensional Bose gas, a quasiparticle excitation of momentum $Q$ decays into three excitations with momenta $q_1,q_2$, and $q_3$. Using the conservation laws, one finds that two quasiparticles in the final state propagate in the direction of the initial quasiparticle, unlike the remaining one.}\label{fig1}
\end{figure}

For one-dimensional particles with the spectrum (\ref{Eq}), decay into two
quasiparticles is incompatible with simultaneous conservation of
energy and momentum.  The simplest allowed decay process corresponds
to three particles in the final state, see Fig.~\ref{fig1}.  It will
become clear below that this is the dominant decay channel in a weakly interacting Bose gas.

We start our evaluation by considering the scattering matrix element
$\mathcal{A}_{fi}$ for the decay of the initial state
$|i\rangle=b_Q^\dagger|0\rangle$ into the final one
$|f\rangle=b_{q_1}^\dagger b_{q_2}^\dagger b_{q_3}^\dagger|0\rangle$.
$\mathcal{A}_{fi}$ is defined in terms of the $T$-matrix as
\begin{align}
\mathcal{A}_{fi} =\langle 0| b_{q_1} b_{q_2}
b_{q_3}|T|b^\dagger_{Q}|0\rangle.
\label{matrix_element}
\end{align}
Such a matrix element can be obtained in a number of ways.  The
simplest contribution is in the first order in the quartic term $V_4$
[Eq.~(\ref{V4})] that allows for the direct transition between the
initial and final states.  Alternatively, the same transition can be
accomplished in second order in the cubic perturbation $V_3$
[Eq.~(\ref{V3})].  In a weakly interacting Bose gas, i.e., at $K\gg1$,
the two perturbations are small, $V_3\propto 1/\sqrt K$ and
$V_4\propto1/K$.  As a result, the two contributions to the matrix
element (\ref{matrix_element}) appear in the same order,
$\mathcal{A}_{fi}\propto1/K$.  A straightforward argument shows that
higher-order anharmonic perturbations to the Hamiltonian $H_0$ give
rise to parametrically smaller contributions to the matrix element
(\ref{matrix_element}).  Accounting only for the leading
contributions, we find
\begin{align}\label{Afi}
\mathcal{A}_{fi}=\langle f|V_4|i\rangle+\sum_m\frac{\langle f|V_3|m\rangle\langle m|V_3|i\rangle} {\varepsilon_Q-E_m}.
\end{align}
Here the summation is over the intermediate states $|m\rangle$, whose
energies are denoted by $E_m$.

The contribution to the scattering matrix element due to the quartic anharmonic term (\ref{V4}) arises from the combinations of operators in $V_4$ that contain three creation and one annihilation operator. There are four such terms. After a simple calculation one obtains
\begin{align}\label{A4full}
\langle f|V_4|i\rangle
={}&\frac{6\pi^2 v^2}{L m n_0} \frac{|Qq_1q_2q_3|}{\sqrt{\varepsilon_{Q} \varepsilon_{q_1} \varepsilon_{q_2} \varepsilon_{q_3}}}\notag\\ &\times f(Q,-q_1,-q_2,-q_3)\delta_{Q,q_1+q_2+q_3},
\end{align}
where the function $f$ is defined in Eq.~(\ref{f3}).

The calculation of the contribution to the scattering matrix element (\ref{Afi}) that arises from $V_3$ is more involved and deferred to Appendix \ref{Appendix:amplitude}. Accounting for Eq.~(\ref{A4full}), the final result for the scattering matrix element (\ref{Afi}) is
\begin{align}\label{Afifinal}
\mathcal{A}_{fi}={}&\frac{\pi^2 v^2}{2Lmn_0} \frac{|Qq_1q_2q_3|}{\sqrt{\varepsilon_Q \varepsilon_{q_1} \varepsilon_{q_2} \varepsilon_{q_3}}} \bigl[F(Q,q_1,q_2,q_3)\notag\\
&+F(Q,q_2,q_1,q_3) +F(Q,q_3,q_2,q_1)\notag\\
&+12f(-Q,q_1,q_2,q_3)\bigr] \delta_{Q,q_1+q_2+q_3},
\end{align}
where we introduced the dimensionless function
\begin{align}\label{Fdef}
F&(q_1,q_2,q_3,q_4)=  \frac{v^2(q_1-q_2)^2}{\varepsilon_{q_1-q_2}}\notag\\
&\times\biggl[ \frac{f_-(q_4, q_3,-q_3-q_4)f_-(q_1-q_2,q_2,-q_1)} {\varepsilon_{q_1}-\varepsilon_{q_2}-\varepsilon_{q_1-q_2}} \notag\\ &-\frac{f_-(q_1,-q_1+q_2,-q_2)f_+(-q_3-q_4,q_3,q_4)} {\varepsilon_{q_3}+\varepsilon_{q_4}+\varepsilon_{q_1-q_2}} \biggr].
\end{align}
Here $\varepsilon_q$ and the functions $f_\pm$ are defined by
Eqs.~(\ref{Eq}) and (\ref{fpm}), respectively.

The scattering matrix element (\ref{Afifinal}) has some important
general properties.  Since $F(q_1,q_2,q_3,q_4)=F(q_1,q_2,q_4,q_3)$,
the matrix element (\ref{Afifinal}) is symmetric with respect to the
exchanges of the momenta of excitations in the final state.  This is a
manifestation of the fact that Bogoliubov quasiparticles obey bosonic
statistics.  More importantly, one can show that at 
\begin{align}\label{LLlimit}
A=B=0,\quad \chi=1
\end{align}
the result (\ref{Afifinal}) vanishes, provided that $q_1$, $q_2$,
$q_3$, and $Q$ satisfy conservation laws of momentum and energy.  This
is because under the conditions (\ref{LLlimit}) our theory describes
the weakly interacting Lieb-Liniger model.  The latter is integrable, and
its quasiparticles do not decay.

We now simplify the scattering matrix element (\ref{Afifinal}) in the
regimes of small and large momenta.

\subsection{Small momentum region}

At small momentum of the initial excitation, $Q\ll {q}_0$, the other
three momenta are also small compared to $q_0$.  In this regime we
have been able to simplify the expression (\ref{Afifinal})
considerably, as discussed in Appendix \ref{appendix:amplitudeexpansion}.
The final result takes the form
\begin{align}\label{AfifinallowQ}
\mathcal{A}_{fi}=\frac{\Lambda}{2Lm n_0}\sqrt{|Qq_1q_2q_3|}\delta_{Q,q_1+q_2+q_3},
\end{align}
where the momentum independent $\Lambda$ is given by
\begin{align}\label{Lambda<general}
\Lambda={}&12\pi^2B-{6\pi^2{a}_1^2}+\frac{24\pi^4{a}_1{a}_2}{\chi^2}\notag\\
&- \frac{{A}}{\pi} \left(18{a}_1+ \frac{24\pi^2{a}_2}{\chi^2}\right).
\end{align}
Using the values of $a_1$, $a_2$, and $\beta\equiv B/K^2$ given by
Eqs.~(\ref{a1a2}) and (\ref{a3}), in the leading order in small
$1-\chi$ we obtain
\begin{align}\label{LambdalowQ}
  \Lambda= -\frac{3\Omega}{\pi^2},
\end{align}
where we defined
\begin{align}\label{Omega}
\Omega=4A-\pi^2(1-\chi).
\end{align} 
We observe again that in the Lieb-Liniger limit (\ref{LLlimit}) the
scattering matrix element vanishes.

\subsection{Large momentum region}

At large momentum of the initial excitation, $Q\gg {q}_0$, we have
also been able to considerably simplify the matrix element
(\ref{Afifinal}).  The main steps are described in Appendix
\ref{appendix:amplitudeexpansion}, resulting in
\begin{gather}\label{AfilargeQ}
\mathcal{A}_{fi}=\frac{2mv^2}{L n_0}\Xi \delta_{Q,q_1+q_2+q_3},
\end{gather}
where
\begin{align}\label{Lambda>}
\Xi={}&12\pi^2B-\frac{23\pi^2{a}_1^2}{8}+\frac{13\pi^4{a}_1{a}_2}{\chi^2}+\frac{26\pi^6{a}_2^2}{\chi^4} \notag\\
&-\frac{4\pi^4{a}_3}{\chi^2} -\frac{A}{\pi}\left(\frac{21}{2}{a}_1 +\frac{30\pi^2{a}_2}{\chi^2}\right).
\end{align}
Substituting the specific values of the parameters of our Hamiltonian
from Eqs.~(\ref{a1a2}) and (\ref{a3}), in the leading order in small
$1-\chi$ we find
\begin{align}
\label{LambdalargeQ}
\Xi=-\frac{9\Omega}{4\pi^2}.
\end{align}
As expected, in the Lieb-Liniger case (\ref{LLlimit}) the scattering
matrix element $\mathcal{A}_{fi}=0$.

\section{Decay rate}\label{sec:decayrate}

Let us now evaluate the rate of decay of a quasiparticle of momentum
$Q>0$ at zero temperature.  The dominant decay process is illustrated in
Fig.~\ref{fig1}.  The corresponding rate of decay is given by the
Fermi golden rule expression
\begin{align}\label{rate def}
\frac{1}{\tau}=\frac{2\pi}{\hbar} \sum_{q_1,q_2,q_3}\!\!\!\!'\,|\mathcal{A}_{fi}|^2 \delta(\varepsilon_{Q}-\varepsilon_{q_1}- \varepsilon_{q_2}-\varepsilon_{q_3}).
\end{align}
The matrix element $\mathcal{A}_{fi}$ describing the decay of the
initial quasiparticle excitation of momentum $Q$ into three
quasiparticles with momenta $q_1$, $q_2$, and $q_3$ is given by
Eq.~(\ref{Afifinal}). The prime symbol in Eq.~(\ref{rate def}) denotes
the summation over distinct final states.

The conservation laws of energy and momentum
\begin{gather}\label{momentumcons}
  Q=q_1+q_2+q_3,\\
  \label{energycons}
  \varepsilon_Q=\varepsilon_{q_1}+ \varepsilon_{q_2}+ \varepsilon_{q_3},
\end{gather}
ensure that out of three new quasiparticles two propagate in the same
direction as the initial quasiparticle, $q_1,q_2>0$, while the
third one is counterpropagating, $q_3<0$, see Fig.~\ref{fig1}.
Conditions (\ref{momentumcons}) and (\ref{energycons}) enable us to
express the momentum of the counterpropagating quasiparticle as a
function of $Q$ and one of the two remaining momenta, for example,
$q_1$. We denote it as $q_3\equiv q_3(Q,q_1)$. With the help of the
two conservation laws we now easily perform two summations in
Eq.~(\ref{rate def}), yielding
\begin{align}\label{rate final uneval}
\frac{1}{\tau}= \frac{L^2}{4\pi\hbar^3} \int_0^{Q}\dif q_1 \frac{|\mathcal{{A}}(Q,q_1,Q-q_1-q_3,q_3)|^2} {|\varepsilon'_{Q-q_1-q_3}-\varepsilon'_{q_3}|},
\end{align}
where
\begin{align}
\varepsilon'_{q}=v\,\text{sgn}(q) \frac{1+4\chi^2\frac{q^2}{{q}_0^2}} {\sqrt{1+2\chi^2\frac{q^2}{{q}_0^2}}}.
\end{align}
In the following we use Eq.~(\ref{rate final uneval}) to evaluate the
quasiparticle decay rate as a function of $Q$.

\subsection{Regime of low momenta}

Let us first consider the case of low momentum of the initial
excitation, $Q\ll{q}_0$, where we recall the definition (\ref{q0}). In
this regime the excitation spectrum is almost linear and thus the
denominator in Eq.~(\ref{rate final uneval}) simplifies into
$2v$. Using the conservation laws (\ref{momentumcons}) and
(\ref{energycons}) we find the leading order result for the momentum
of the counterpropagating excitation
\begin{align}\label{q3small}
q_3= -\frac{3Qq_1}{2{q}_0^2}(Q-q_1).
\end{align}
Substituting it in Eq.~(\ref{rate final uneval}) with the matrix element given by (\ref{AfifinallowQ}), after integration we obtain
\begin{align}\label{ratefinallowQgeneral}
\frac{1}{\tau}=\frac{9\sqrt{2}}{5\pi} \frac{\Omega^2}{K^4} \frac{T_d}{\hbar} \left(\frac{Q}{{q}_0}\right)^7.
\end{align}
Here we introduced the quantum degeneracy temperature $T_d=\hbar^2
n_0^2/m$.

In the limit of contact interaction, $\chi=1$, and the decay rate (\ref{ratefinallowQgeneral}) becomes
\begin{align}\label{ratefinallowQ}
\frac{1}{\tau}=\frac{144\sqrt{2}}{5\pi} \alpha^2\frac{T_d}{\hbar} \left(\frac{Q}{{q}_0}\right)^7.
\end{align}
This result was found earlier in Ref.~\cite{ristivojevic_decay_2014}
using a phenomenological approach, in which the phonon is treated as a
mobile impurity.  Here we rederived that result fully microscopically
and generalized it to the case of noncontact interaction.

\subsection{Regime of high momenta}

Now we consider the case of large momentum of the initial excitation, $Q\gg{q}_0$. The conservation laws (\ref{momentumcons}) and (\ref{energycons}) can be easily solved when all quasiparticles are in the quadratic part of the spectrum. One finds
\begin{gather}\label{eq:q2bigQ}
q_2=\frac{1}{2}\left[Q-q_1+\sqrt{(Q-q_1)(Q+3q_1)}\right],\\
\label{eq:q3bigQ}
q_3=\frac{1}{2}\left[Q-q_1-\sqrt{(Q-q_1)(Q+3q_1)}\right].
\end{gather}
The latter expressions enable us to simplify the denominator 
in Eq.~(\ref{rate final uneval}), which becomes $2\sqrt{2}v\sqrt{(Q-q_1)(Q+3q_1)}/q_0$. Here we take into account the leading order result in small $1-\chi$. Using the matrix element (\ref{AfilargeQ}) and the expression
\begin{align}
\int_0^Q \frac{\dif q_1}{\sqrt{(Q-q_1)(Q+3q_1)}}=\frac{2\sqrt{3}\pi}{9},
\end{align}
we obtain the decay rate of quasiparticles of large momenta:
\begin{align}\label{ratefinalhighQgeneral}
\frac{1}{\tau}=\frac{9\sqrt{3}}{8} \frac{\Omega^2}{K^4} \frac{T_d}{\hbar}.
\end{align}
We note that the expression (\ref{rate final uneval}) contains regions
of integration where the momentum $q_1$ is either close to zero or
$Q$.  In these regions two quasiparticles of the final state are in
the linear part of the spectrum, where the approximations
(\ref{eq:q2bigQ}) and (\ref{eq:q3bigQ}) fail.  We checked that the
contributions arising from these boundary regions give only a
subleading correction to the decay rate (\ref{ratefinalhighQgeneral}).

In the limit of contact interaction $\chi=1$. 
Equation (\ref{ratefinalhighQgeneral}) then reduces to 
\begin{align}\label{ratefinalhighQtan}
\frac{1}{\tau}=18\sqrt{3}\alpha^2\frac{T_d}{\hbar}.
\end{align}
This result was obtained earlier in Ref.~\cite{tan+10relaxation} using a different approach.

\subsection{The crossover regime}

\begin{figure}
\includegraphics[width=0.95\columnwidth]{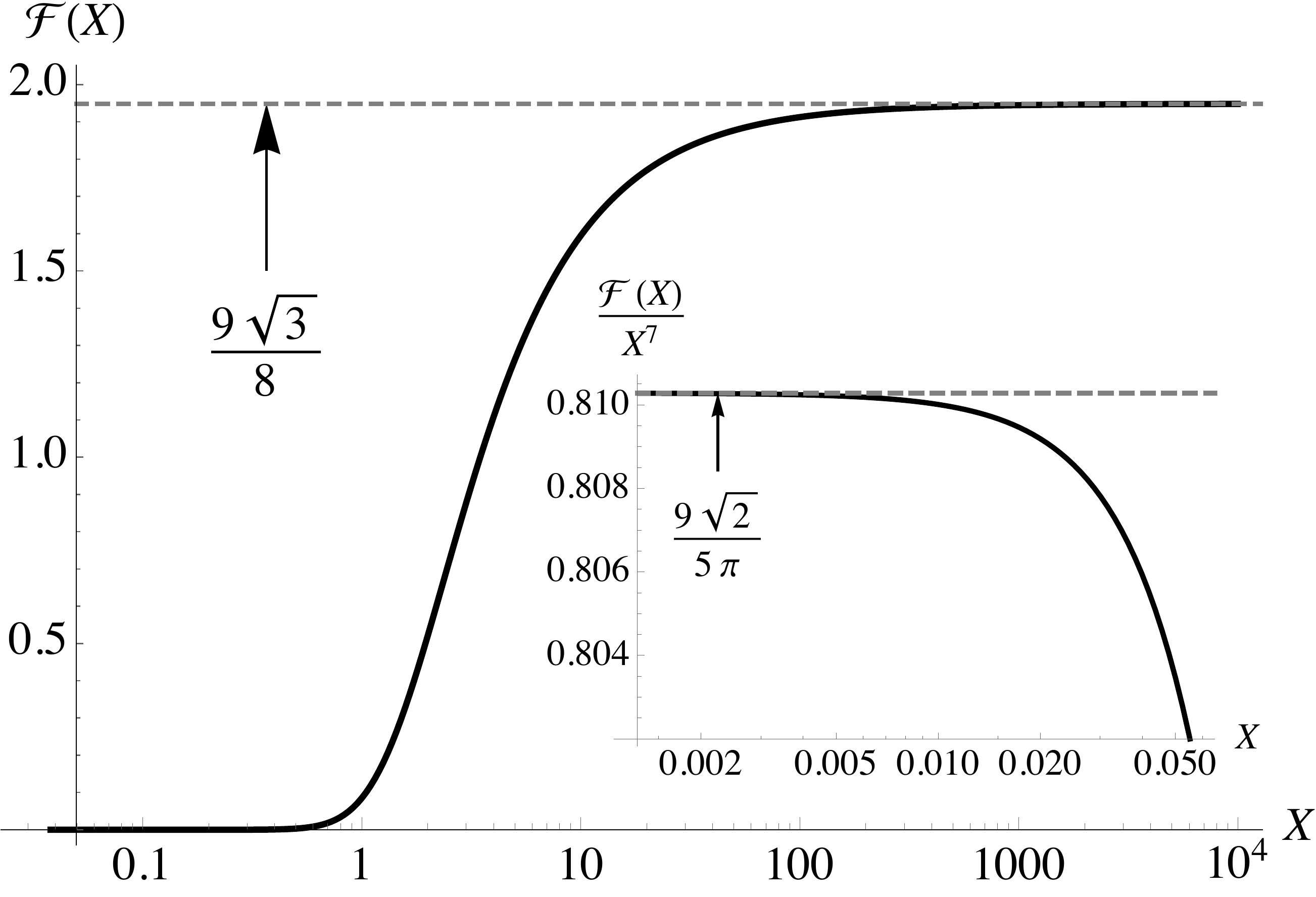}
\caption{Plot of the function $\mathcal{F}(X)$ given by Eqs.~(\ref{E17})-(\ref{E20}) that enters the relaxation rate (\ref{rate-general}). The inset shows the limiting behavior of $\mathcal{F}(X)$ at $X\to 0$.}\label{fig3crossover}
\end{figure}

In the regime of intermediate momenta, $Q\sim {q}_0$, complete analytical evaluation of the decay rate (\ref{rate final uneval})
is a challenging problem. However, we are able to express it in the form 
\begin{align}\label{rate-general}
\frac{1}{\tau}=\frac{\Omega^2}{K^4} \frac{T_d}{\hbar}\mathcal{F}\left(\frac{Q}{q_0}\right),
\end{align}
where $\Omega$ is given by Eq.~(\ref{Omega}).  The analytical form for
the function $\mathcal{F}$ is given by Eqs.~(\ref{E17})-(\ref{E20}) of
Appendix \ref{section:lagrange}. It has the asymptotic behavior
\begin{align}\label{FF}
\mathcal{F}(X)=\begin{cases}\frac{9\sqrt{2}}{5\pi}X^7,\quad &X\ll 1,\\
\frac{9\sqrt{3}}{8},\quad &X\gg 1.
\end{cases}
\end{align}
The latter result is in agreement with already calculated decay rates in the limiting cases of low [Eq.~(\ref{ratefinallowQgeneral})] and high [Eq.~(\ref{ratefinalhighQgeneral})] momenta. In Fig.~\ref{fig3crossover} we plot the function $\mathcal{F}$. 

\section{Decay of fermionic excitations at low energies}\label{sec:fermions}

Description of elementary excitations of weakly interacting Bose gas
in terms of phonons with Bogoliubov dispersion (\ref{Eq}) is
applicable only at sufficiently high momenta.  Indeed, the correction
to the linear spectrum $\varepsilon_q=v|q|$ in Eq.~(\ref{Eq}) is due to
the term proportional $(\nabla^2\varphi)^2$ in the Hamiltonian
(\ref{H0}).  At $q\to0$ the relative significance of a perturbation in
the Hamiltonian is determined by its scaling dimension, which for the
operator $(\nabla^2\varphi)^2$ is four.  On the other hand,
perturbations $\nabla \varphi (\nabla\theta)^2$ and
$(\nabla\varphi)^3$ of lower scaling dimension three are also present
in the Hamiltonian, see Eq.~(\ref{V3general}).  At the lowest
energies, the latter perturbations control the physics of the
elementary excitations and their spectrum \cite{rozhkov2005fermionic}.
Specifically, the excitations at $q\to0$ are fermions with spectrum
\begin{equation}
  \label{eq:fermion_spectrum}
  \varepsilon_q=v|q|+\frac{q^2}{2m^*}+\frac16\lambda^* |q|^3+\ldots.
\end{equation}
Most importantly, unlike the Bogoliubov dispersion (\ref{Eq}), the
leading correction is quadratic, with finite effective mass $m^*$.  

To determine $m^*$ and $\lambda^*$ it is sufficient to consider the
low-momentum part of the hydrodynamic Hamiltonian, accounting for the
right-moving excitations only.  This is accomplished by substituting
\begin{eqnarray}
  \label{eq:phi_chiral}
  \varphi&=&\frac{\sqrt K}{2} (\phi^L+\phi^R),
\\
  \label{eq:theta_chiral}
  \theta&=&\frac{1}{2\sqrt K} (\phi^L-\phi^R)
\end{eqnarray}
into Eqs.~(\ref{H0}), (\ref{V3general}), and (\ref{V4general}) and
limiting oneself to terms containing only the right-moving field
$\phi^R$.  The leading operator of this form 
\begin{equation}
  \label{eq:tilde_H_LL}
  \tilde H_{LL}=\frac{\hbar v}{4\pi}\int dx \left(\nabla \phi^R\right)^2
\end{equation}
is simply the right-moving part of the Luttinger liquid Hamiltonian
(\ref{HLL}).  It has scaling dimension two and is responsible for the
linear part of the excitation spectrum in
Eq.~(\ref{eq:fermion_spectrum}).  The terms of scaling dimensions
three and four can be combined into
\begin{equation}
  \label{eq:H_KdV}
  H_{\rm KdV}=\frac{\hbar^2}{12\pi m^*}\int dx 
         \left[\left(\nabla \phi^R\right)^3 
          + a^*\left(\nabla^2 \phi^R\right)^2\right],
\end{equation}
where
\begin{eqnarray}
  \label{eq:effective_mass}
  \frac{1}{m^*}&=&\frac{1}{m}\,
                \frac{3}{4\sqrt K}\left(1-\frac{2}{\pi^2}A\right),
\\
  a^*&=&\frac{\hbar\chi^2 \sqrt K}{2mv}\left(1-\frac{2}{\pi^2}A\right)^{-1}.
\end{eqnarray}
The Hamiltonian (\ref{eq:H_KdV}) describes one of the possible
realizations of the quantum KdV problem \cite{sasaki_field_1987,
  pogrebkov_boson-fermion_2003}.  The spectrum of elementary
excitations in this model has been recently studied in detail in
Ref.~\cite{pustilnik_fate_2015}.  At $q\to0$ it has Taylor expansion
(\ref{eq:fermion_spectrum}) with $\lambda^*=\chi^2/4m^2v$.  The
crossover from fermionic excitations to phonons with Bogoliubov
dispersion occurs at momentum scale $q^*\sim\hbar/a^*\sim q_0/\sqrt
K\ll q_0$.

At $Q\ll q^*$ type I and type II excitations (see Fig.~\ref{figLL})
correspond to fermionic quasiparticles and quasiholes, respectively.
In the absence of integrability, quasiparticles can decay at zero
temperature, with the rate that scales as the eighth power of momentum
\cite{khodas+07PhysRevB.76.155402,matveevfurusaki13},
\begin{equation}
  \label{eq:fermionic_decay}
  \frac1\tau=\frac{3}{5120\pi^3}\,
             \frac{\tilde\Lambda^2Q^8}{\hbar^5m^* v^2}.
\end{equation}
A general expression for the coefficient $\tilde\Lambda$ in terms of
the parameters $v$, $m^*$ and $\lambda^*$ was obtained in
Ref.~\cite{matveevfurusaki13}.  At weak interactions, $K\gg 1$, the
expression for $\tilde\Lambda$ simplifies significantly,
\begin{equation}
  \label{eq:tilde_Lambda}
  \tilde\Lambda=-\frac{2\pi}{3m^*}
                \frac{\partial}{\partial n_0}\left(a^*\sqrt K\right).
\end{equation}
This result was recently obtained for a one-dimensional Wigner crystal
\cite{pustilnik_solitons_2015}, whose low-energy excitations are also
described by the Hamiltonian in the form of Eqs.~(\ref{eq:tilde_H_LL})
and (\ref{eq:H_KdV}).

In the integrable case of the Lieb-Liniger model achieved at $A=0$ and
$\chi=1$ one easily sees that $a^*\sqrt K$ does not depend on particle
density $n_0$, and the decay rate vanishes.  Taking into consideration
the integrability breaking perturbations described by parameters $A$
and $1-\chi$ that both scale linearly with $n_0$ [see Eqs.~(\ref{A})
and (\ref{chi2})], we obtain
\begin{equation}
  \label{eq:tilde_Lambda_result}
  \tilde\Lambda=-\frac{2\hbar^2\Omega}{3m^*m^2v^2}.
\end{equation}
Substituting this expression into Eq.~(\ref{eq:fermionic_decay}) we
find the decay rate of the fermionic quasiparticle in the form
\begin{equation}
  \label{eq:fermionic_decay_result}
  \frac{1}{\tau}=\frac{9}{20\pi}\,
                 \frac{\Omega^2}{K^{7/2}}\,
                 \frac{T_d}{\hbar}\,
                 \left(\frac{Q}{q_0}\right)^8.
\end{equation}
Reassuringly, at the crossover between Bogoliubov phonons and
fermions, i.e., at $Q\sim q_0/\sqrt K$, both the expressions
(\ref{ratefinallowQgeneral}) and (\ref{eq:fermionic_decay_result})
predict a very small rate $\tau^{-1}\sim\Omega^2(T_d/\hbar)K^{-15/2}$.

\section{Discussion}\label{sec:discussions}

In this paper we studied the decay of type~I excitations in
a one-dimensional system of weakly interacting bosons at zero
temperature.  The approach we used was based on the hydrodynamic
description of the system, which limits the momenta of the bosons to
$Q\ll\hbar n_0$.  Two additional momentum scales play important roles
in this system.  First, the momentum $q_0=\sqrt8mv\sim \hbar n_0/K$
determines the crossover between the linear and quadratic dependences
of the excitation energy (\ref{Eq}) on momentum.  Second, at the
momentum scale $q^*\sim q_0/\sqrt K$ the nature of type~I excitations
changes from fermionic quasiparticles at $Q\ll q^*$ to phonons at
$Q\gg q^*$.  We note that at weak interactions the Luttinger liquid
parameter $K\gg1$, thus $q^*\ll q_0\ll \hbar n_0$.

Our main result (\ref{rate-general}) applies in the region $q^*\ll
Q\ll\hbar n_0$ and accurately describes the crossover region $Q\sim
q_0$.  In addition, we obtained the decay rate of the fermionic
quasiparticles at $Q\ll q^*$.  Although we are not able to describe
the crossover at $Q\sim q^*$, our results (\ref{ratefinallowQgeneral})
and (\ref{eq:fermionic_decay_result}) for $Q\gg q^*$ and $Q\ll q^*$,
respectively, give the decay rate of the same order of magnitude when
extrapolated to $Q\sim q^*$.  This strongly indicates that no
additional crossover regions remain unexplored.

It is instructive to compare our result (\ref{rate-general}) to those
in the earlier work on weakly interacting bosons.  In the case of
contact two-body repulsion the system is described by the Lieb-Liniger
model, in which case the integrability prevents decay of excitations.
A small perturbation commonly added to the system to break
integrability is the three-body interaction given by the second term
in Eq.~(\ref{Hint}).  In this case the regimes $Q\gg q_0$ and $q^*\ll
Q\ll q_0$ were studied in Refs.~\cite{tan+10relaxation} and
\cite{ristivojevic_decay_2014}, respectively.  Our main result
(\ref{rate-general}) recovers the corresponding expressions
(\ref{ratefinalhighQtan}) and (\ref{ratefinallowQ}) for the decay rate
and accurately describes the crossover between them.

An alternative way to break the integrability of the Lieb-Liniger
model is by considering two-body interaction of small but finite
range.  Our theory incorporates this perturbation on equal footing
with the three-body interactions.  The relative significance of the
two perturbations depends on the specific model of interacting bosons.
In the case of atoms confined to one dimension by a trap, we expect
the three-body interaction to dominate \cite{ristivojevic+matveev-unpublished}.  On
the other hand, noncontact interactions in a purely one-dimensional
model should generate the three-body interactions in the effective
low-energy theory, in which case both perturbations may be of the same
order of magnitude.  

To illustrate this point, we have considered the hyperbolic
Calogero-Sutherland model in the regime of weak short-range
interaction.  It is defined by the two body interaction of the form
\cite{sutherland}
\begin{align}\label{sinh22}
g(x)=\frac{\hbar^2}{m} \frac{\lambda(\lambda-1)\kappa^2}
{\sinh^2(\kappa x)}.
\end{align}
In the limit when $\kappa\to+\infty$ and $\lambda\to+0$, such that
$c=2\kappa\lambda$ is kept fixed, the scattering matrix of the
potential (\ref{sinh22}) coincides with that of the potential
$g(x)=(\hbar^2 c/m)\delta(x)$ \cite{sutherland}.  Therefore, in this
limit the model (\ref{sinh22}) is equivalent to the Lieb-Liniger
model.  We then obtained the excitation spectrum of the model
(\ref{sinh22}) at large but finite $\kappa$, see Appendix
\ref{sec:calogero-sutherland}.  Using the latter, we have found the
values of parameters $\alpha$ and $\chi$ that quantify the two
integrability-breaking perturbations:
\begin{align}
\alpha=-\frac{\pi^2c^2}{24\kappa^2},\quad \chi=1+\frac{\pi^2 cn_0}{6\kappa^2}.
\end{align}
We observe that for the integrable model (\ref{sinh22}), the
combination (\ref{Omega}) becomes
\begin{align}\label{univratio}
\Omega=4K^2\alpha-\pi^2(1-\chi)=0.
\end{align}
We therefore conclude that the two perturbations give comparable
contributions to the scattering amplitude corresponding to the decay
process, which for the model (\ref{sinh22}) cancel each other.  This
cancellation was, of course, expected, as the hyperbolic
Calogero-Sutherland model (\ref{sinh22}) is integrable for any
$\kappa$ and $\lambda$ \cite{sutherland}.

\section*{Acknowledgements}

We acknowledge stimulating discussions with L.~I.~Glazman and 
M.~Pustilnik. K.~A.~M. is grateful to Laboratoire de Physique Th\'{e}orique,
Toulouse, where part of the work was performed, for hospitality.  Work
by K.~A.~M.~was supported by the U.S.~Department of Energy, Office of
Science, Materials Sciences and Engineering Division.

\appendix

\section{Scaling analysis of the hydrodynamic Hamiltonian}
\label{scaling}

Our main goal is to study the decay rate of Bogoliubov quasiparticles
at momenta of the order of the crossover value $q_0\sim mv$, assuming
that the interactions are weak.  The latter condition can be expressed
as $q_0\ll \hbar n_0$ or $K\gg1$, cf. Eq.~(\ref{K}).  To this end we
apply the following procedure to the hydrodynamic Hamiltonian given by
Eqs.~(\ref{H0}), (\ref{V3general}), and (\ref{V4general}).

We rescale the lengths by the scale determined by $q_0$, i.e., introduce
\begin{gather}
\widetilde x=x q_0/\hbar.
\label{x-rescaled}
\end{gather}
Correspondingly, the derivative transforms as
\begin{gather}
\nabla=\frac{q_0}{\hbar}\widetilde\nabla.
\end{gather}
At the same time, we rescale the bosonic fields according to
\begin{gather}
\varphi=\sqrt{K}\,\widetilde\varphi,
\quad 
\theta=\frac{\widetilde\theta}{\sqrt{K}}.
\label{fields-rescaled}
\end{gather}
Note that the above procedure preserves the commutation relations
of the bosonic fields,
$[\widetilde\nabla\widetilde\varphi(\widetilde
x),\widetilde\theta(\widetilde x')]=-i\pi\delta(\widetilde
x-\widetilde x')$.  In rescaled variables the contributions
(\ref{H0}), (\ref{V3general}), and (\ref{V4general}) to the
Hamiltonian become
\begin{subequations}\label{Hresc}
\begin{gather}
{H}_0=\frac{vq_0}{2\pi}\int\dif\widetilde x \left[(\widetilde\nabla \widetilde\theta)^2 + (\widetilde\nabla \widetilde\varphi)^2 +2 (\widetilde\nabla^2 \widetilde\varphi)^2\right],\\
{V}_3=\frac{\sqrt{2} vq_0}{\pi} \frac{1}{\sqrt{K}} \int\dif \widetilde x \biggl[ (\widetilde\nabla \widetilde\varphi) (\widetilde\nabla \widetilde\theta)^2-2 (\widetilde\nabla^2 \widetilde\varphi)^2(\widetilde\nabla \widetilde\varphi)\notag\\
-\frac{2A}{\pi^2}(\widetilde\nabla \widetilde\varphi)^3\biggr],
\\
{V}_4=\frac{8vq_0}{\pi}\frac{1}{K}
  \int\dif \widetilde x\left[ (\widetilde\nabla^2 \widetilde\varphi)^2 
  (\widetilde\nabla \widetilde\varphi)^2
  +\pi^2B  (\widetilde\nabla \widetilde\varphi)^4
   \right].
\label{V4resc}
\end{gather}
\end{subequations}
Here we introduced
\begin{align}\label{Atilde}
B=K^2 \beta
\end{align}
and substituted the values (\ref{a1a2}) and (\ref{a3}) of the
constants $a_1$, $a_2$, and $a_3$.  

The scaling procedure (\ref{x-rescaled})--(\ref{fields-rescaled})
enables one to estimate the relative significance of the various
contributions to the Hamiltonian describing the physics of the system
at the momentum scale $q_0$ and the respective energy scale $vq_0$.
Contributions $H_0$, $V_3$, and $V_4$ represent the first three terms
of the expansion of the Hamiltonian in small parameter $1/\sqrt K$.
The terms of higher orders in the bosonic fields continue this trend.
Indeed, all such terms emerge from the expansion of the density $n$ in
the denominator of the quantum pressure term in Eq.~(\ref{Hkin}) with
the help of Eq.~(\ref{density}).  Rewriting the latter expression in
rescaled variables, we obtain
\begin{equation}
  \label{eq:1}
  n=n_0\left(
       1+\sqrt\frac{8}{K}\,\widetilde\nabla\widetilde\varphi
       \right).
\end{equation}
Thus each additional order in the bosonic field $\widetilde\varphi$ is
accompanied by a small coefficient of order $1/\sqrt K$.

\section{Second order perturbation theory for the scattering matrix element}\label{Appendix:amplitude}

In this appendix we present some details of the evaluation of the matrix element defined by Eq.~(\ref{Afi}). We first consider the contribution arising from the cubic perturbation $V_3$ of Eq.~(\ref{V3}). Using the identity
\begin{align}
\frac{1}{E+i\delta}=\frac{1}{i\hbar}\int_0^\infty\dif t e^{i t (E+i\delta)/\hbar},\quad \delta>0,
\end{align}
we perform the summation over the intermediate states,  reexpressing the scattering matrix element that arises due to $V_3$ as
\begin{align}\label{summation}
\sum_m\frac{\langle f|V_3|m\rangle \langle m|V_3|i\rangle} {\varepsilon_Q-E_m+i\delta}
  =\int_0^\infty \dif t \frac{e^{-t\delta/\hbar}}{i\hbar} \langle f|V_3(0) V_3(-t)|i\rangle.
\end{align}
Here we use the operators in the Heisenberg representation $V_3(t)=e^{i t H_0/\hbar}V_3e^{-i t H_0/\hbar}$,
where $H_0$ is the quadratic part of the Hamiltonian, see Eqs.~(\ref{H0}) and (\ref{H0diag}). The creation and annihilation operators in this picture are
\begin{align}
b_q(t)=e^{-it\varepsilon_q/\hbar}b_q,\quad
b_q^\dagger(t)=e^{it\varepsilon_q/\hbar}b_q^\dagger.
\end{align}

Direct inspection on Eq.~(\ref{V3}) reveals that out of sixteen possible terms, only three of them may give nonzero contribution in $\langle f|V_3(0) V_3(-t)|i\rangle$, because they contain an equal number of creation and annihilation operators. One of them nullifies after performing Wick contractions due to the momentum conservation, while the remaining terms are
\begin{widetext}
\begin{align}\label{matrixelement}
\langle  f|V_3(0) V_3(-t)|i\rangle=& \frac{\pi^2 v^4}{8Lm n_0} \sum_{p_1,p_2,p_3\atop p_1',p_2',p_3'}\delta_{p_1+p_2+p_3,0} \delta_{p_1'+p_2'+p_3',0} \frac{|p_1p_2p_3p_1'p_2'p_3'|} {\sqrt{\varepsilon_{p_1} \varepsilon_{p_2} \varepsilon_{p_3} \varepsilon_{p_1'} \varepsilon_{p_2'} \varepsilon_{p_3'}}}\notag\\
&\times\biggl[\frac{1}{3}f_-(p_1,p_2,p_3)f_+(p_1',p_2',p_3')\big\langle b_{q_3}b_{q_2}b_{q_1} b_{-p_3}^\dagger b_{p_2} b_{p_1} b_{p_1'}^\dagger b_{p_2'}^\dagger b_{p_3'}^\dagger b_{Q}^\dagger\big\rangle e^{-it(\varepsilon_{p_1'} +\varepsilon_{p_2'}+ \varepsilon_{p_3'})/\hbar}\notag\\
&+f_-(p_1,p_2,p_3) f_-(p_1',p_2',p_3')\big\langle b_{q_3}b_{q_2}b_{q_1}  b_{p_1}^\dagger b_{p_2}^\dagger b_{-p_3} b_{p_1'}^\dagger b_{p_2'}^\dagger b_{-p_3'}b_{Q}^\dagger\big\rangle e^{-it(\varepsilon_{p_1'} +\varepsilon_{p_2'}- \varepsilon_{p_3'})/\hbar}\biggr].
\end{align}

\end{widetext}

We now use Wick theorem \cite{wickPhysRev.80.268,tyablikov} to evaluate the expression (\ref{matrixelement}). Denoting \begin{align}
C_1=\big\langle b_{q_3}b_{q_2}b_{q_1} b_{-p_3}^\dagger b_{p_2} b_{p_1} b_{p_1'}^\dagger b_{p_2'}^\dagger b_{p_3'}^\dagger b_{Q}^\dagger\big\rangle
\end{align}
we note that $b_Q^\dagger$ must not be contracted with any of $b_{q_1}$, $b_{q_2}$, or $b_{q_3}$ because in this case the energy and momentum conservation would imply zero value for the remaining two momenta in the final state $|f\rangle$, which is not the scattering process we consider. Thus, we contract $b_Q^\dagger$ with, for example, $b_{p_1}$ and account for a factor of 2 because $f_-(p_1,p_2,p_3)=f_-(p_2,p_1,p_3)$. Then the other operator $b_{p_2}$ must be contracted with either of $b_{p_1'}^\dagger$, $b_{p_2'}^\dagger$, or $b_{p_3'}^\dagger$. Because $f_+(p_1',p_2',p_3')$ is fully symmetric with respect to the permutations of its arguments, we arbitrary select, e.g., $b_{p_1'}^\dagger$ and account for a factor of 3 in this choice. We then obtain
$
C_1=6\delta_{Q,p_1}\delta_{p_2,p_1'} \big\langle b_{q_3}b_{q_2}b_{q_1} b_{-p_3}^\dagger b_{p_2'}^\dagger b_{p_3'}^\dagger \big\rangle.
$
We note that the last expression is symmetric with respect to the permutations of $q_1$, $q_2$, and $q_3$. Therefore, the six remaining contractions we write in a compact notation introducing the symmetrization operator $\mathcal{\hat S}$ that denotes the summation over all permutations of $q_1$, $q_2$, and $q_3$: \begin{align}
C_1=6\delta_{Q,p_1}\delta_{p_2,p_1'} \mathcal{\hat S}(\delta_{-p_3,q_1} \delta_{p_2',q_2}\delta_{p_3',q_3}).
\end{align}
We note that there will actually be only three different terms in the final result for the matrix element, since the momenta $p_2'$ and $p_3'$ under the symmetrization operator in the last expression enter symmetrically because $f_+(p_1',p_2',p_3')$ is already symmetric, see Eq.~(\ref{matrixelement}).

The other combination of the operators in Eq.~(\ref{matrixelement}) we denote by
\begin{align}
C_2=\big\langle b_{q_3}b_{q_2}b_{q_1}  b_{p_1}^\dagger b_{p_2}^\dagger b_{-p_3} b_{p_1'}^\dagger b_{p_2'}^\dagger b_{-p_3'}b_{Q}^\dagger\big\rangle.
\end{align}
Clearly, $b_{Q}^\dagger$ must be contracted with $b_{-p_3'}$. Next, $b_{-p_3}$  must be contracted with either $b_{p_1'}^\dagger$ or $b_{p_2'}^\dagger$. We select $b_{p_1'}^\dagger$ and account for a factor of 2, which yields
$
C_2=2\delta_{Q,-p_3'}\delta_{-p_3,p_1'}\big\langle b_{q_3}b_{q_2}b_{q_1}  b_{p_1}^\dagger b_{p_2}^\dagger b_{p_2'}^\dagger\big\rangle.
$
The last expression we rewrite using the symmetrization operator
\begin{align}
C_2=2\delta_{Q,-p_3'}\delta_{-p_3,p_1'} \mathcal{\hat S}(\delta_{p_1,q_1} \delta_{p_2,q_2}\delta_{p_2',q_3}).
\end{align}
Similarly as in $C_1$, in $C_2$ we eventually have only three distinct contributions since $f_-(p_1,p_2,p_3)$ is symmetric with respect to the interchange of $p_1$ and $p_2$.

Substituting those results and performing the summation, the matrix element (\ref{matrixelement}) becomes
\begin{widetext}
\begin{align}
\langle  f|V_3(0) V_3(-t)|i\rangle=&\frac{\pi^2 v^4}{4Lmn_0}\frac{|Qq_1 q_2 q_3 q_4|}{\sqrt{\varepsilon_Q \varepsilon_{q_1} \varepsilon_{q_2} \varepsilon_{q_3 }}} \mathcal{\hat S}{\biggl (}\frac{(q_2+q_3)^2}{\varepsilon_{q_2+q_3}} f_-(Q,q_1-Q,-q_1)f_+(-q_2-q_3,q_2,q_3) e^{-it(\varepsilon_{q_2+q_3}+\varepsilon_{q_2} +\varepsilon_{q_3})/\hbar}\notag\\
&+ \frac{(Q-q_3)^2}{\varepsilon_{Q-q_3}} f_-(q_1,q_2,-q_1-q_2)f_-(Q-q_3,q_3,-Q) e^{-it(\varepsilon_{Q-q_3}+\varepsilon_{q_3} -\varepsilon_{Q})/\hbar}\biggr) \delta_{Q,q_1+q_2+q_3}.
\end{align}
\end{widetext}
After performing the integration over $t$, see Eq.~(\ref{summation}), in the limit $\delta\to +0$, we obtain the final result contained in the expression (\ref{Afifinal}).

\section{Evaluation of the scattering matrix element}

\label{appendix:amplitudeexpansion}

In this Appendix we provide some details of the evaluation of the scattering matrix element (\ref{Afifinal}) in the regimes of small and large momenta. Due to the conservation laws, we select the momenta to satisfy $Q>q_1,q_2>0$ and $q_3<0$, see Fig.~\ref{fig1}.

The function $F$ that determines the amplitude is defined by Eq.~(\ref{Fdef}).  We conveniently split it as $F=F_1-F_2$, where 
\begin{align}
F_1&(q_1,q_2,q_3,q_4)={}\frac{v^2(q_1-q_2)^2}{\varepsilon_{q_1-q_2}}\notag\\
&\times\frac{f_-(q_4, q_3,-q_3-q_4)f_-(q_1-q_2,q_2,-q_1)} {\varepsilon_{q_1}-\varepsilon_{q_2}-\varepsilon_{q_1-q_2}},\\
F_2&(q_1,q_2,q_3,q_4)=\frac{v^2(q_1-q_2)^2}{\varepsilon_{q_1-q_2}}\notag\\
&\times\frac{f_-(q_1,-q_1+q_2,-q_2)f_+(-q_3-q_4,q_3,q_4)} {\varepsilon_{q_3}+\varepsilon_{q_4}+\varepsilon_{q_1-q_2}}.
\end{align}

\subsection{Small momentum region}
\label{Appendix:amplitudesmallQ}

At small momenta, $Q,q_1,q_2,|q_3|\ll q_0$, in ${F}_2$ terms we can safely linearize the spectrum at low momenta, i.e., we can use $\varepsilon_q=v|q|$. It yields
\begin{align}
{F}_2(Q,q_1,q_2,q_3)={}&{F}_2(Q,q_2,q_1,q_3)= {F}_2(Q,q_3,q_2,q_1)=\notag\\& \frac{1}{2} \left({a}_1+\frac{3{A}}{\pi^3}\right)^2 +\ldots,
\end{align}
where the ellipsis denotes the subleading terms that tend to zero at small momenta.

In the $F_1$ part we have to keep the spectrum nonlinearity because of the energy denominator. We use the momentum conservation to replace $Q$ by $q_1+q_2+q_3$, and then we use the expression for the smallest momentum given by 
\begin{align}\label{q3q1q2}
q_3=-\frac{3q_1q_2}{2q_0^2}(q_1+q_2).
\end{align}
Finally, we expand the obtained expression at small $q_1,q_2\ll q_0$, keeping the ratio $q_1/q_2$ fixed. We obtain
\begin{align}\label{F1eval}
F_1(Q,q_1,q_2,q_3)=& -{{a}_1^2}\frac{6\hat{q}_0^2 +13q_1^2+22q_1q_2+7q_2^2} {6q_1(q_1+q_2)}\notag\\
&+
\frac{8\pi^2{a}_1{a}_2 }{\chi^2} \frac{q_1^2+q_1q_2-2q_2^2}{3q_1(q_1+q_2)}\notag\\
&-\frac{{a}_1{A}}{\pi^3} \frac{6\hat{q}_0^2+17q_1^2+26q_1q_2+11q_2^2} {3q_1(q_1+q_2)}\notag\\
&+\frac{{A}^2}{\pi^6} \frac{6\hat{q}_0^2+ 5q_1^2+14q_1q_2-q_2^2}{2q_1(q_1+q_2)}\notag\\
&+\frac{8{a}_2 {A}}{\pi\chi^2}
\frac{q_1^2+q_1q_2+2q_2^2}
{q_1(q_1+q_2)}+\ldots,
\end{align}
where for notational convenience we introduced $\hat q_0=q_0/\chi$. The terms in the ellipsis contain the subleading terms. 

The expression for $F_1(Q,q_2,q_1,q_3)$ is trivially obtained from Eq.~(\ref{F1eval}) by exchanging momenta $q_1$ and $q_2$, both of which assumed to be positive.

The remaining term $ F_1(Q,q_3,q_2,q_1)$ cannot be directly inferred from Eq.~(\ref{F1eval}) because $q_3$ enters the expression in a special way, and is negative by the initial assumption. By repeating the above procedure and using the conservation laws to rewrite $1/(\varepsilon_Q-\varepsilon_{q_3}-\varepsilon_{Q-q_3})$ as $1/(\varepsilon_{q_1}+\varepsilon_{q_2}-\varepsilon_{q_1+q_2})$ and again using Eq.~(\ref{q3q1q2}) to remove $q_3$, we eventually expand at small $q_1$ and $q_2$. We obtain
\begin{align}\label{F1eval1}
F_1(Q,q_3,q_2,q_1)=& {{a}_1^2}\frac{6\hat{q}_0^2 +7q_1^2+q_1q_2+7q_2^2} {6q_1q_2}\notag\\
&+
\frac{8\pi^2{a}_1{a}_2 }{\chi^2} \frac{2q_1^2+5q_1q_2+2q_2^2}{3q_1q_2}\notag\\
&+\frac{{a}_1{A}}{\pi^3} \frac{6\hat{q}_0^2+11q_1^2+5q_1q_2+11q_2^2} {3q_1q_2}\notag\\
&+\frac{{A}^2}{\pi^6} \frac{-6\hat{q}_0^2 +q_1^2+7q_1q_2+q_2^2}{2q_1q_2}\notag\\
&-\frac{8{a}_2 {A}}{\pi\chi^2}
\frac{2q_1^2+3q_1q_2+2q_2^2}
{q_1q_2}+\ldots
\end{align}
Alternatively, if we do not use the conservation law to transform the energy denominator, we need to find the subleading correction in the result for $q_3$ [Eq.~(\ref{q3q1q2})], which is a more involved, but an equivalent way to obtain Eq.~(\ref{F1eval1}). As expected, Eq.~(\ref{F1eval1}) is symmetric with respect to the exchange of momenta $q_1$ and $q_2$.

When we sum all the terms, recalling $F=F_1-F_2$, we obtain the leading contribution at low momenta
\begin{align}\label{Fsum}
{F}&(Q,q_1,q_2,q_3)+{F}(Q,q_2,q_1,q_3)+ {F}(Q,q_3,q_2,q_1)\notag\\
=&-{6{a}_1^2}+\frac{24\pi^2{a}_1{a}_2}{\chi^2}- \frac{{A}}{\pi^3} \left(18{a}_1+ \frac{24\pi^2{a}_2}{\chi^2}\right).
\end{align}

\subsection{Large momentum region}
\label{Appendix:amplitudelargeQ}

In the regime of large momentum of the initial excitation, $Q\gg q_0$, at least one momentum of quasiparticles in the final state is of the same order as $Q$. Let us denote it by $q_1$. At such high momenta, the Bogoliubov dispersion (\ref{Eq}) can be approximated as
$\varepsilon_q =\chi q^2/2m + m v^2/\chi$. Using the latter expression one can easily solve the conservation laws of the momentum and energy
to find $q_2$ and $q_3$:
\begin{align}\label{q2bigQ}
q_2={}&\frac{1}{2}\left[Q-q_1+\sqrt{(Q-q_1)(Q+3q_1)}\right] \notag\\
&-\frac{2m^2v^2}{\chi^2\sqrt{(Q-q_1)(Q+3q_1)}},\\
\label{q3bigQ}
q_3={}&\frac{1}{2}\left[Q-q_1-\sqrt{(Q-q_1)(Q+3q_1)}\right]\notag\\
&+\frac{2m^2v^2}{\chi^2\sqrt{(Q-q_1)(Q+3q_1)}}.
\end{align}
Substituting the latter expressions in Eq.~(\ref{Afifinal}),
after some algebra one obtains the final result given by Eqs.~(\ref{AfilargeQ}) and Eq.~(\ref{Lambda>}).
Let us comment that in Eq.~(\ref{Lambda>}) we neglected small terms that are momentum dependent. The leading one in that expansion is proportional to $[(1-\chi^2)Q/\chi q_0]^2$. Since $1-\chi$ is small, the latter term imposes the condition 
\begin{gather}
Q\ll \frac{q_0}{\sqrt{1-\chi}}.
\end{gather}
It does not affect the decay rate in a wide region around $q_0$, since $1-\chi$ is the small parameter.

\section{Hyperbolic Calogero-Sutherland model}\label{sec:calogero-sutherland}

In Sec.~\ref{sec:amplitude} we found the general expression for the
scattering matrix element for the decay process of a Bogoliubov
quasiparticle shown in Fig.~\ref{fig1}. At low momenta, it is given by
Eq.~(\ref{AfifinallowQ}), while the expression at high momenta is
Eq.~(\ref{AfilargeQ}).  In this Appendix we take advantage of
integrability of the hyperbolic Calogero-Sutherland model to find the
parameters $\alpha$ [cf.~Eq.~(\ref{V3general})] and $\chi$
[cf.~Eqs.~(\ref{H0}) and (\ref{Eq})] that enter the prefactors of the
scattering matrix element.

We consider the two-body interaction potential for the hyperbolic Calogero-Sutherland model \cite{sutherland}
\begin{align}\label{sinh2}
g(x)=\frac{\hbar^2}{m} \frac{\lambda(\lambda-1)\kappa^2}
{\sinh^2(\kappa x)}.
\end{align}
A many-body problem of bosons interacting with the potential (\ref{sinh2}) is exactly solvable by Bethe ansatz. An important quantity for this technique is the two-particle scattering phase shift, which is given by \cite{sutherland} 
\begin{align}\label{thetaplus} 
\theta_+(k)&=i\log\left(\frac{\Gamma(1+ik/2\kappa) \Gamma(\lambda-ik/2\kappa)}{ \Gamma(1-ik/2\kappa) 
\Gamma(\lambda+ik/2\kappa)}\right).
\end{align}
This complicated function has an important yet simple limiting case. Namely, at $\kappa\to+\infty$, $\lambda\to +0$, such that $c=2\kappa\lambda$ is kept fixed, the phase shift (\ref{thetaplus}) coincides with the one of the Lieb-Liniger model \cite{sutherland}:
\begin{align}\label{thetaLL}
\theta_+(k)\underset{\kappa\to+\infty\atop \lambda\to +0}{\longrightarrow} \theta_{\text{LL}}(k)=-2\arctan\left(\frac{k}{c}\right).
\end{align}
The phase shift (\ref{thetaLL}) corresponds to the interaction potential $g(x)=g\delta(x)$, where  $c=m g/\hbar^2$. 

The relation between the two integrable models enables us to consider
corrections to the Lieb-Liniger model caused by finite interaction
range without losing the integrability.  To this end, we account for
the leading deviation in Eq.~(\ref{thetaLL}) due to large but finite
$\kappa$.  While Eq.~(\ref{thetaLL}) is valid at any $c$, here we are
interested in the limit of weak interaction. Therefore, we expand
$\theta_+(k)-\theta_{\text{LL}}(k)$ in linear order at small $c$ and
obtain
\begin{align}\label{theta+exp}
\theta_+(k)=\theta_{\text{LL}}(k) +\frac{\pi c}{2\kappa}
\left[\coth\left(\frac{\pi k}{2\kappa}
\right)-\frac{2\kappa}{\pi k}\right].
\end{align}
The correction terms in this expression account for the deviation of the Calogero-Sutherland model from the Lieb-Liniger model at weak interaction due to large but finite $\kappa$. The phase shift (\ref{theta+exp}) contains necessary information to find the excitation spectrum of the hyperbolic Calogero-Sutherland model in this particular limit of small interaction range.

At small wavevectors, $k\ll \kappa$, Eq.~(\ref{theta+exp}) contains a nontrivial correction to the phase shift of the Lieb-Liniger model. The leading order expression in the large-$\kappa$ limit is
\begin{align}\label{theta+fin}
\theta_+(k)&=-2\arctan \left(\frac{k}{c}\right) +\frac{\pi^2 ck}{12\kappa^2}.
\end{align}
The knowledge of the phase shift (\ref{theta+fin}) suffices to find the excitation spectrum at not too high momenta. It is parametrically given by
\begin{align}\label{pkek}
p(k)=2\pi\hbar\int_{k_0}^k \dif q \rho(q),\quad \varepsilon(k)=\int_{k_0}^k \dif q \sigma(q),
\end{align}  
where $k_0$ is the Fermi rapidity. We consider particle-like excitations, so we study the case $k>k_0$. In the previous equation $\rho(k)$ is the density of rapidities. It is determined by the Lieb's integral equation
\begin{align}\label{ILeq}
\rho(k)+\frac{1}{2\pi}\int_{-k_0}^{+k_0}\dif 
q \theta_+'(k-q)\rho(q)=\frac{1}{2\pi},
\end{align}
and normalized as
\begin{align}\label{norm}
\int_{-k_0}^{k_0}\dif q\rho(q)=n_0.
\end{align}
The other density function $\sigma(k)$ in Eq.~(\ref{pkek}) satisfies a similar equation
\begin{align}\label{IILeq}
\sigma(k)+\frac{1}{2\pi}\int_{-k_0}^{+k_0}\dif 
q \theta_+'(k-q)\sigma(q)=\frac{\hbar^2 k}{m}.
\end{align}

The two integral equations can be treated in the limit of large $\kappa$ by iterations. Their solution at $k>k_0$ can be expressed as
\begin{gather}
\rho(k)=\left(1-\frac{\pi^2 k_0^2}{48\kappa^2}\right)\frac{1}{2\pi}
\frac{\dif}{\dif k}\sqrt{k^2-k_0^2},
\end{gather}
\begin{gather}
\sigma(k)=\frac{\hbar^2}{6m}
\frac{\dif^2}{\dif k^2}(k^2-k_0^2)^{3/2}.
\end{gather}
Substituting them in Eq.~(\ref{pkek}) we find the spectrum
\begin{gather}\label{disp}
\varepsilon_p=\sqrt{\frac{\hbar^2k_0^2 p^2}{4m^2\mathcal{N}^2}+\frac{p^4}{4m^2\mathcal{N}^4}},\quad \mathcal{N}=1-\frac{\pi^2k_0^2}{48\kappa^2}.
\end{gather}
The normalization condition (\ref{norm}) leads to the Fermi rapidity 
\begin{align}\label{k0}
k_0=2\sqrt{c n_0}\left(1+\frac{\pi^2cn_0}{24\kappa^2}\right).
\end{align}
This result requires the knowledge of $\rho(k)$ function at momenta below $k_0$: $\rho(k)=\mathcal{N}\sqrt{k_0^2-k^2}/2\pi c$. The density functions below and above $k_0$ are connected by the Lieb's integral equation.

Substituting the value of $k_0$ in Eq.~(\ref{disp}) we easily obtain linear spectrum at low momenta, $\varepsilon_p=v |p|$, where 
\begin{align}\label{vCS}
v=\frac{\hbar}{m}\sqrt{cn_0} 
\left(1+\frac{\pi^2 cn_0}{8\kappa^2}\right).
\end{align}
The latter expression is the sound velocity of the hyperbolic Calogero-Sutherland model at large $\kappa$. As expected, at $\kappa\to+\infty$ one obtains the familiar expression for the sound velocity of the Lieb-Liniger model \cite{lieb1963exact}. We notice that the sound velocity can be also found from the thermodynamic expression 
\begin{align}
v=\sqrt{\frac{L}{m n_0}\frac{\partial^2 E_0}{\partial L^2}},
\end{align}
where
\begin{align}
E_0=\frac{\hbar^2 L}{2m}\int_{-k_0}^{k_0}\dif k\, k^2\rho(k)
\end{align}
denotes the ground state energy.

Detailed knowledge of the sound velocity as a function of the density is sufficient to find the parameters $\alpha$
[cf.~Eq.~(\ref{V3general})] and $\beta$ [cf.~Eq.~(\ref{V4general})] in
the hydrodynamic Hamiltonian.  They can be expressed as \cite{landaubook-9}
\begin{align}
\alpha=-\frac{m^2}{6\hbar^2}\frac{\dif }{\dif n_0}\left(\frac{v^2}{n_0}\right),\quad \beta=\frac{m^2n_0}{24\pi^4\hbar^2} \frac{\dif^2 }{\dif n_0^2}\left(\frac{v^2}{n_0}\right).
\end{align}
Substituting the velocity (\ref{vCS}) in the latter expression enables us to find at $1/\kappa^2$ order
\begin{align}
\label{eq:alpha_result}
\alpha=-\frac{\pi^2c^2}{24\kappa^2},\quad \beta=0.
\end{align}

The expression (\ref{disp}) for the energy of excitations obtained in
leading order in $1/\kappa$ has the form (\ref{Eq}).  This enables us
to obtain the expression for the parameter $\chi$ in the hyperbolic
Calogero-Sutherland model
\begin{align}\label{chispec}
\chi=1+\frac{\pi^2 
cn_0}{6\kappa^2}.
\end{align}
Comparing Eqs.~(\ref{eq:alpha_result}) and (\ref{chispec}) we find the following relation between $A=K^2\alpha$ and $\chi$,
\begin{align}
\frac{A}{1-\chi}=\frac{\pi^2}{4}.
\end{align}
When the latter condition is satisfied and $\beta=0$, the matrix
element (\ref{Afifinal}) nullifies at all momenta.  This observation
is consistent with the expected absence of decay of excitations in
integrable models.

We note that the parameter $\chi$ could also be obtained from the
definitions (\ref{chi2}) and (\ref{sinh2}).  Formally, Fourier
transform of the potential (\ref{sinh2}) diverges. However, the
second derivative at zero momentum, which enters Eq.~(\ref{chi2}), is
finite:
\begin{align}
\frac{\dif^2 g_q}{\dif q^2}\bigg{|}_{q=0}
=-\frac{1}{\hbar^2} \int_{-\infty}
^{+\infty}\dif x x^2 g(x)= -\frac{\pi^2 
\lambda(\lambda-1)}{3m\kappa}.
\end{align}
The limiting procedure $\kappa\to+\infty$, $\lambda\to +0$ with fixed
$c=2\kappa\lambda$ reproduces Eq.~(\ref{chispec}).

\section{Transformation of the matrix element using Lagrangian
  variables}
\label{section:lagrange}

\subsection{Lagrangian description of interacting one-dimensional
  bosons}
\label{sec:Lagrangian}

Our derivation of the hydrodynamic Hamiltonian in
Sec.~\ref{density-phase} was based on Eqs.~(\ref{madelung}) and
(\ref{density}) that replace the bosonic operators of the particles
constituting the Bose gas with new bosonic fields $\varphi(x)$ and
$\theta(x)$, which describe the state of the fluid at point $x$.
Alternatively, one can consider a uniform reference state of density
$n_0$ and develop hydrodynamics in terms of the fields that are
functions of the coordinate $y$ of the fluid element in that state
\cite{matveev+12PhysRevB.86.045136}.  This approach corresponds to
using Lagrangian variables in the classical fluid dynamics, as opposed
to the standard Eulerian ones \cite{landaubook-6}.  The physical position $x$ of
the fluid element is obtained by adding its displacement $u$ to the
position $y$ in the reference state,
\begin{equation}
  \label{eq:x_vs_y}
  x=y+u(y).
\end{equation}
Using Eq.~(\ref{eq:x_vs_y}) one obtains the expression for the
particle density
\begin{equation}
  \label{eq:density_Lagrangian}
  n(y)=\frac{n_0}{1+u'(y)},
\end{equation}
where prime denotes the derivative with respect to $y$.  Comparing
this expression with Eq.~(\ref{density}) we obtain the relation
\begin{equation}
  \label{eq:varphi_vs_u}
  \frac{1}{\pi}\nabla\varphi(x)=-\frac{u'(y)}{1+u'(y)}
\end{equation}
between the fields $\varphi(x)$ and $u(y)$.

In addition to the displacement $u(y)$, one can introduce the momentum
density $p(y)=mn_0v(y)$ of the fluid.  The two fields obey the
standard commutation relation
\begin{equation}
  \label{eq:commutator_u_p}
  [u(y),p(y')]=i\hbar\delta(y-y').
\end{equation}
A relation between $p(y)$ and the Eulerian field $\theta(x)$ is found
by comparing the above definition of $p(y)$ with the expression for
the velocity of the fluid $v(x)=-(\hbar/m)\nabla\theta(x)$, see Ref.~\cite{matveev+12PhysRevB.86.045136}.  This yields
\begin{equation}
  \label{eq:theta_vs_p}
  \nabla\theta(x)=-\frac{1}{\hbar n_0}\,p(y).
\end{equation}
One can now substitute Eqs.~(\ref{eq:x_vs_y}), (\ref{eq:varphi_vs_u}),
and (\ref{eq:theta_vs_p}) into the hydrodynamic Hamiltonian
approximated by Eqs.~(\ref{H0}), (\ref{V3general}), and
(\ref{V4general}) and obtain an equivalent theory in Lagrangian
variables.

For our needs it is helpful to first rescale the new bosonic fields
$u$ and $p$ as follows:
\begin{eqnarray}
  \label{eq:Phi}
  u(y)&=&-\frac{1}{\pi n_0}\, \Phi(y),
\\
  p(y)&=&-\hbar n_0\nabla\Theta(y)
  \label{eq:Theta},
\end{eqnarray}
where $\nabla=d/dy$.  The new fields satisfy the commutation relation
\begin{equation}
  \label{eq:commutator_Phi_Theta}
  [\Phi(y),\nabla\Theta(y')]=i\pi\delta(y-y').
\end{equation}
The convenience of the bosonic fields $\Phi$ and $\Theta$ manifests
itself in the equivalence of the commutation relations
(\ref{eq:commutator_phi_theta}) and (\ref{eq:commutator_Phi_Theta})
upon replacing
\begin{equation}
  \label{eq:replacement}
  x\to y,
\quad
  \varphi\to\Phi,
\quad
  \theta\to\Theta.
\end{equation}
Combining the Eqs.~(\ref{eq:x_vs_y}), (\ref{eq:varphi_vs_u}), (\ref{eq:theta_vs_p}), (\ref{eq:Phi}), and (\ref{eq:Theta}) we obtain the following
relations between the bosonic fields in Eulerian and Lagrangian
variables:
\begin{subequations}
  \label{eq:transformation_to_Lagrange}
\begin{eqnarray}
  x&=&y-\frac{1}{\pi n_0}\Phi(y),
\\
  \nabla\varphi(x)&=&\frac{\nabla\Phi(y)}{1-(1/\pi n_0)\nabla\Phi(y)},
\\
  \nabla\theta(x)&=&\nabla\Theta(y).
\end{eqnarray}
\end{subequations}
One should stress that $\nabla$ here denotes the derivative with
respect to the appropriate variable, i.e., $\nabla=d/dx$ and $d/dy$ in
the left- and right-hand sides of
Eq.~(\ref{eq:transformation_to_Lagrange}), respectively.

Substituting Eq.~(\ref{eq:transformation_to_Lagrange}) into the
hydrodynamic Hamiltonian given by Eqs.~(\ref{H0}), (\ref{V3general}),
and (\ref{V4general}), we find that, up to contributions of higher
than quartic order in the bosonic fields, the Hamiltonian retains its
general form, provided that the parameters
$\{a_1,a_2,a_3,\alpha,\beta\}$ are replaced with
$\{a_1^L,a_2^L,a_3^L,\alpha^L,\beta^L\}$ given by
\begin{subequations}\label{Lagrangeparameters}
\begin{eqnarray}
a_{1}^L&=&a_1-\frac{1}{2\pi},\\
a_2^L&=&a_2-\frac{5\chi^2}{8\pi^3},\\
a_3^L&=&a_3-\frac{6a_2}{\pi}+\frac{15\chi^2}{8\pi^4},\\
\alpha^L&=&\alpha-\frac{\pi^2}{2K^2},\\
\beta^L&=&\beta-\frac{2\alpha}{\pi^4}+\frac{1}{2\pi^2 K^2}.
\end{eqnarray}
\end{subequations}
Because the physics of the system cannot be sensitive to our choice of
Eulerian or Lagrangian variables used to describe it, all the
physically observable quantities should be invariant with respect to
the change of parameters from $\{a_1,a_2,a_3,\alpha,\beta\}$ to
$\{a_1^L,a_2^L,a_3^L,\alpha^L,\beta^L\}$.  In particular, it is easy
to check that  the expressions for $\Lambda$ and $\Xi$ given,
respectively, by Eqs.~(\ref{Lambda<general}) and (\ref{Lambda>}) are
invariant with respect to this transformation.  

Using the values of the constants given by Eqs.~(\ref{a1a2}) and
(\ref{a3}), we eventually obtain
\begin{subequations}
\label{Lagrangeconstants}
\begin{align}
\label{eq:a_1_Lagrangian}
a_{1}^L={}&0,\\
a_2^L={}&\frac{1-5\chi^2}{8\pi^3},\\
a_3^L={}&-\frac{5}{8\pi^4}(1-3\chi^2),\\
A^L={}&A-\frac{\pi^2}{2},\\
B^L={}&-\frac{2A}{\pi^4}+\frac{1}{2\pi^2}.
\label{eq:B_Lagrangian}
\end{align}
\end{subequations}
Use of Lagrangian variables greatly simplifies the following
calculations.

\subsection{Evaluation of the scattering martix element
  (\ref{Afifinal})}
\label{sec:matrix_element}

We now apply the hydrodynamic theory in terms of Lagrangian variables
to the evaluation of the matrix element (\ref{Afifinal}) responsible
for the decay of Bogoliubov excitations.  The main advantage of using
the Lagrangian variables is that the coefficient $a_1$ entering the
definitions (\ref{fpm}) of functions $f_\pm$ now vanishes, see
Eq.~(\ref{eq:a_1_Lagrangian}).  One should keep in mind that in this
case the constant $B$ takes a nonvanishing value
(\ref{eq:B_Lagrangian}), which somewhat complicates the expression
(\ref{f3}) for the function $f$.

We have been able to show that to first order in $A$ and $1-\chi$ the matrix element (\ref{Afifinal}) nullifies for any set of the four momenta that satisfy the conservation laws (\ref{momentumcons}) and (\ref{energycons}), provided that
\begin{align}
\Omega\equiv 4A-\pi^2(1-\chi)=0.
\end{align}
This observation enables us to simplify Eq.~(\ref{Afifinal}) using a
fixed value $\chi=1$ by collecting terms linear in $A$.  This
calculation benefits greatly from using Lagrangian variables.  Upon
restoring nonzero $1-\chi$ in the final expression, we obtain
\begin{align}\label{AfiLagrange}
\mathcal{A}_{fi}&=-\frac{3\Omega v^2}{8 \pi^2 Lmn_0}\frac{|Qq_1q_2q_3|}{\sqrt{\varepsilon_{Q} \varepsilon_{q_1} \varepsilon_{q_2} \varepsilon_{q_3}}}\bigl[8+ f_L(Q,q_1,q_2,q_3)\notag\\
&+f_L(Q,q_2,q_1,q_3)+f_L(Q,q_3,q_1,q_2)\bigr]\delta_{Q,q_1+q_2+q_3},
\end{align}
where
\begin{align}
f_L(Q,q_1,q_2,q_3)={}&m^2\biggl[\frac{(\varepsilon_Q+\varepsilon_{q_1})^2-\varepsilon_{Q-q_1}^2}{Q^2 q_1^2}\notag\\
&+\frac{(\varepsilon_{q_2}-\varepsilon_{q_3})^2-\varepsilon_{q_2+q_3}^2}{q_2^2 q_3^2}\biggr].
\end{align}
The latter expression assumes the dispersion $\varepsilon_q=\sqrt{v^2 q^2+q^4/4m^2}$, i.e., one should replace $\chi=1$ in Eq.~(\ref{Eq}). The expression in brackets in Eq.~(\ref{AfiLagrange}) interpolates between $4$ at $Q\ll mv$ and $3$ at $Q\gg mv$. We thus recover the results (\ref{AfifinallowQ}) and (\ref{AfilargeQ}) for $B=0$.

The decay rate (\ref{rate final uneval}) can be conveniently evaluated using Eq.~(\ref{AfiLagrange}). After introducing the dimensionless momenta $X=Q/q_0$, $x=q_1/q_0$, $y=q_2/q_0$, and $z=q_3/q_0$, the decay rate assumes the form (\ref{rate-general}), with the crossover function given by
\begin{widetext}
\begin{align}\label{E17}
\mathcal{F}(X)={}&\frac{9\sqrt{2}X^2}{128\pi\epsilon_X}\int_{0}^{X} \dif x \frac{x^2y^2z^2}{\epsilon_{x} \epsilon_{y} \epsilon_{z}}\biggl[64+
\frac{(\epsilon_X+\epsilon_x)^2-\epsilon_{X-x}^2}{X^2x^2} +\frac{(\epsilon_X+\epsilon_y)^2-\epsilon_{X-y}^2}{X^2y^2}+ \frac{(\epsilon_X+\epsilon_z)^2-\epsilon_{X-z}^2}{X^2z^2}\notag\\
&+\frac{(\epsilon_y-\epsilon_z)^2-\epsilon_{y+z}^2}{y^2z^2} +\frac{(\epsilon_x-\epsilon_z)^2-\epsilon_{x+z}^2}{x^2z^2}+ \frac{(\epsilon_x-\epsilon_y)^2-\epsilon_{x+y}^2}{x^2y^2}\biggr]^2 \left(\frac{1+4y^2}{\sqrt{1+2y^2}}+\frac{1+4z^2}{\sqrt{1+2z^2}}\right)^{-1}.
\end{align}
\end{widetext}
In this formula we introduced the dimensionless energy as
$\epsilon_x=\sqrt{x^2+2x^4}$, while the dimensionless momentum
$y=X-x-z$ is fixed by the momentum conservation. Finally, $z$ is
obtained as a negative solution of the energy conservation equation
that takes the form
$\epsilon_X=\epsilon_x+\epsilon_{X-x-z}+\epsilon_z$. It can be
expressed as
\begin{align}\label{z}
z=\frac{X-x}{2}-\frac{1}{4}\sqrt{R_1+\sqrt{R_2}},
\end{align}
where
\begin{align}
R_1={}&\left(\frac{\epsilon_X - \epsilon_x}{X-x}\right)^2 - 
2  (2+3 X^2 + 3 x^2 - 2 X x),\\
\label{E20}
R_2={}&\left(\frac{\epsilon_X - \epsilon_x}{X-x}\right)^4 +4\left(\frac{\epsilon_X - \epsilon_x}{X-x}\right)^2(2+11X^2+11x^2
\notag\\ 
&-18Xx)-4(X+x)^2(4+7X^2+7x^2-2Xx).
\end{align}	
At small momenta, $X,x\ll 1$, from Eq.~(\ref{z}) we recover the expression (\ref{q3small}). Similarly, at $X,x\gg 1$, Eq.~(\ref{z}) leads to the result (\ref{q3bigQ}).

%

\end{document}